\documentclass[
 reprint,
 superscriptaddress,
 nofootinbib,
 floatfix,
 amsmath,amssymb,
 aps,
 prc
]{revtex4-2}

\usepackage{graphicx}
\usepackage{bm}
\usepackage{booktabs}
\usepackage{dcolumn}
\usepackage{xcolor}
\usepackage[colorlinks=true,linkcolor=blue,citecolor=blue,urlcolor=blue]{hyperref}

\newcommand{\nuc}[2]{\ensuremath{{}^{#1}\mathrm{#2}}}

\newcommand{\fm}{\ensuremath{\mathrm{fm}}}
\newcommand{\dif}{\mathrm{d}}

\begin{document}

\title{Mapping parametric error profiles onto nuclear structure configurations in deformed proton radioactivity}

\author{Jizheng Bo}
\email[]{bojzh22@tongji.edu.cn}
\affiliation{School of Physics Science and Engineering, Tongji University, Shanghai 200092, China}

\date{\today}

\begin{abstract}
Theoretical descriptions of proton radioactivity near the drip lines are often challenged by parametric uncertainties in nuclear potential models. To address this, a robust Bayesian uncertainty quantification (UQ) framework is established to calibrate the deformed Woods--Saxon potential coupled with the semi-classical WKB approximation for odd-$A$ proton emitters across the $Z=50$--$82$ region. Constrained by experimental half-lives, the joint and marginal posteriors and covariance topologies are extracted for the potential geometry, multipole deformations, and spectroscopic factors. Using a signed relative sensitivity index, a dynamic, mass-dependent re-ordering of parameter hierarchies is unraveled. Crucially, propagating these parametric uncertainties yields posterior predictive medians and differentiated $\sigma$ bands that tightly encompass experimental half-lives across multiple orders of magnitude without systematic bias. Furthermore, near the $Z=82$ shell closure, the structural softening of the nuclear potential manifests as a systematic expansion and volatility of the $\sigma$ bands in the deformation profiles. This correlation suggests that the macroscopic behavior of parametric error distributions can potentially be mapped onto potential energy configurations, underscoring that the Bayesian UQ methodology may serve as a sensitive probe to extract reliable nuclear structure information directly from deformed proton decay systematics.
\end{abstract}

\maketitle

\section{Introduction}
Proton radioactivity, as a fundamental quantum-mechanical tunneling process, serves as a powerful spectroscopic tool for exploring the structural evolution and stability limits of exotic nuclei near the proton drip line\cite{RevModPhys.84.567,PhysRevLett.131.202501}. Since its discovery in 1970\cite{JACKSON1970281}, this decay mode has provided direct access to single-particle energies, spectroscopic factors, and orbital angular momenta of unbound nuclear states\cite{DELION2006113,PhysRevC.93.034325}. In the transitional mass region between the proton magic numbers $Z=50$ and $Z=82$\cite{RevModPhys.77.427,PhysRev.75.1969,PhysRev.75.1766.2, hinke2012superallowed}, the interplay between core polarization\cite{Wang_2025,MAHAUX19851} and single-particle degrees of freedom\cite{PhysRevLett.95.232502,nilsson1955binding,PhysRevC.89.014319,RevModPhys.75.121} triggers rich structural phenomena, such as shape coexistence\cite{nfgm-wmg2}, collective rotations\cite{PhysRevLett.86.5866}, and rapid onset of nuclear deformations\cite{RevModPhys.82.2155}. Consequently, describing proton emission in this region demands a theoretical framework that accurately accounts for the orientation-dependent barrier geometry modulated by multipole deformations.

Over the past few decades, a variety of theoretical models have been successfully developed to describe the proton-decay dynamics and calculate half-lives. These primarily include semi-empirical and phenomenological formulas, such as the Geiger--Nuttall-like law\cite{QI2019214,PhysRevLett.96.072501} and modified Viola--Seaborg plots\cite{PhysRevC.87.024308} tailored for proton emission. On a more microscopic level, frameworks combining relativistic or non-relativistic mean-field (RMF)\cite{RING1996193,MENG2006470,PhysRevC.5.626,PhysRevC.21.1568} approaches with the $R$-matrix theory\cite{Descouvemont_2010,DESCOUVEMONT2016199} or cluster configurations\cite{RevModPhys.90.035004,PhysRevC.94.024315,VONOERTZEN200643} have been widely employed to study the single-particle structure of drip-line nuclei. Concurrently, macro-microscopic models like the Generalized Liquid Drop Model (GLDM)\cite{PhysRevC.79.054330} and density-dependent folding models\cite{PhysRevC.84.064307,PhysRevC.105.024327,PhysRevC.83.014310} utilizing effective nucleon-nucleon interactions (e.g., DDM3Y)\cite{PhysRevC.105.014612,BHATTACHARYA2007263,PhysRevC.72.051601} have achieved notable success. For strongly deformed non-spherical proton emitters, advanced coupled-channels\cite{PhysRevC.81.024315,Delion_2018} calculations and the semi-classical Wentzel--Kramers--Brillouin (WKB)\cite{PhysRevC.101.054310,PhysRevC.76.027303,PhysRevC.68.034319} approximation combined with a deformed Woods--Saxon potential\cite{PhysRevC.88.064327} have established themselves as reliable and computationally efficient paradigms. 

Despite the success of these diverse models, traditional implementations of the deformed WS-potential framework frequently suffer from intrinsic limitations in parameter optimization and predictive robustness. Typically, the numerous input parameters—including the potential depth, bulk radius, diffuseness, spectroscopic factor, and multipole deformations—are determined via deterministic global fits or adopted directly from macro-microscopic mass models. Such approaches often mask strong parameter correlations and degeneracies, particularly within soft transitional regimes where multiple deformation configurations\cite{PhysRevC.63.014315,RevModPhys.83.1467} may yield equally passable half-life predictions. More importantly, standard deterministic calculations provide only single point-value predictions, lacking the capacity to evaluate statistical confidence bounds and propagate epistemic uncertainties from experimental inputs to final half-life outputs. 

To overcome these paradigms, the incorporation of Bayesian uncertainty quantification (UQ)\cite{Phillips_2021,8h93-j8y7,PhysRevC.111.034005,JAISWAL2026140243} has emerged as a revolutionary trend in modern nuclear physics, offering a rigorous statistical pipeline for parameter calibration and correlation analysis. By translating parameter constraints into continuous joint posterior probability distributions via Markov-chain Monte Carlo (MCMC) sampling\cite{van2018simple,brooks2011handbook,berg2004introductionmarkovchainmonte}, the Bayesian framework enables a comprehensive, multi-dimensional diagnostic of theoretical models rather than a superficial point estimation. While Bayesian UQ has been successfully implemented in nuclear mass assessments and charge radii systematics, its application to full-scale parameter propagation and systematic sensitivity evaluations within the deformed proton-decay framework remains desirable. In the present work, this methodology is deployed across the $Z=50$--$82$ region for odd-$A$ proton emitters to extract the full marginalized posteriors and parameter covariances. Through a signed relative sensitivity index, a dynamic re-ordering of parameter hierarchies along the $Z$ sequence is unraveled. Crucially, the calculated posterior median predictions, paired with their differentiated $\sigma$ uncertainty bands, demonstrate high explanatory performance and robust agreement with experimental half-lives across multiple orders of magnitude.

A notable aspect of the deformed potential model is the presence of parameter degeneracies, often referred to as sloppiness\cite{PhysRevLett.97.150601,transtrum2015perspective} in the Bayesian posterior landscape. This phenomenon, where significant variations in multiple parameters can compensate each other while yielding similar half-life predictions, is particularly pronounced in certain transitional nuclei such as \nuc{113}{Cs} (see Appendix for details). Quantifying such sloppiness is crucial for assessing the reliability and intrinsic limitations of theoretical predictions for proton radioactivity.

The remainder of this paper is structured as follows. Section~\ref{sec:theory} details the theoretical formulation of the deformed potential, the multipole Coulomb expansion, the orientation-dependent WKB approximation, and the Bayesian formulation. Section~\ref{sec:calc} presents the posterior distributions, the systematic evolution of potential parameters and global sensitivities along the $Z$ sequence, and the corresponding uncertainty quantification of half-lives. Finally, conclusions and perspectives for future extensions are provided in Sec.~\ref{sec:conclusion}.

\section{Theoretical framework}
\label{sec:theory}
In this section, the Bayesian formulas and proton decay theory involving deformed potential will be introduced.

\subsection{Bayesian formulation}

The core of a Bayesian framework is to acquire the probability distribution of the quantities of interest, i.e., poseterior function.

The posterior takes the form
\begin{equation}
  p(\bm{\alpha}|\mathbf{D})=
  \frac{p(\mathbf{D}|\bm{\alpha})p(\bm{\alpha})}{p(\mathbf{D})}
  \propto p(\mathbf{D}|\bm{\alpha})p(\bm{\alpha}),
  \label{eq:bayes}
\end{equation}
where $p(\mathbf{X}|\mathbf{Y})$ is the probability of $\mathbf{X}$ given $\mathbf{Y}$. Consequently, $\mathbf{D}$ refers to experiment data, $\bm{\alpha}$ refers to the parameter vector. Then $p(\mathbf{D}|\bm{\alpha})$ represents likelihood function. And $p(\bm{\alpha})$ denotes the prior, with $p(\mathbf{D})$ being marginal distribution of the data.

Concretely, a posterior $p(\bm{\alpha}|\bm{D})$ describes the probability that the parameters (or the models) are correct after seeing the data. Also, a prior $p(\bm{\alpha})$ explains what is known before seeing the data. Moreover, a likelihood $p(\mathbf{D}|\bm{\alpha})$ provides the information about how well the parameters (models) describes the data. Lastly, the denominator $p(\mathbf{D})$ serves as a marginal distribution of the data given the likelihood and prior function, which often in effect is 
regarded as some normalization constant in practical calculations.

One can usually obtain a specific likelihood in the form below. Let $\bm{\alpha}=(\alpha_1, \alpha_2, \ldots, \alpha_n)$ denote the parameter vector containing all model parameters to be calibrated. The likelihood function for the full dataset is constructed as a product of likelihoods for individual measurements,

\begin{equation}
  p(\mathbf{D}|\bm{\alpha})\propto
  \prod_i p(D_i|\bm{\alpha})=
  \exp\left[
  -\frac{1}{2}\sum_{i}
  \frac{(y_i-f(x_i,\bm{\alpha}))^2}
  {\sigma_{i}^2}
  \right],
  \label{eq:likelihood}
\end{equation}
where the sum runs over all data points $i$, with $y_i$ being the experimental observable and $f(x_i,\bm{\alpha})$ being the theoretical prediction depending on the parameter vector $\bm{\alpha}$. The quantity $\sigma_{i}$ represents the experimental uncertainty for nucleus $i$.

Independent priors are assigned to the model parameters
\begin{equation}
  p(\bm{\alpha})\propto
  \exp\left[
  -\frac{1}{2}\sum_i
  \frac{(\alpha_i-\bar{\alpha}_i)^2}{\sigma_{\alpha_i}^2}
  \right],
  \label{eq:prior}
\end{equation}
where the hyperparameters $\{\sigma_{\alpha_i}\}$ and $\{\bar{\alpha}_i\} $ describe the variance and mean of those parameters of interest respectively.

\subsection{Proton-decay potential}

The proton-emission system has been constructed as a proton and a deformed daughter nucleus, the state of proton $j^\pi$\cite{DELION2006113} should be obtained by
\begin{equation}
  |I_P-I_D|\le j\le I_P+I_D,\qquad
  \pi_P=(-1)^l\pi_D,
\end{equation}
where $I_P$ and $I_D$ are the spins of the parent and daughter nuclei, $\pi_P$ and $\pi_D$ are the parities of the parent and daughter nuclei, and $j$ is the total angular momentum of the emitted proton.

The effective proton-core potential is taken as
\begin{equation}
  V(r,\theta)=V_N(r,\theta)+V_C(r,\theta)+V_{so}(r,\theta)
  +\frac{\hbar^2}{2\mu r^2}\left(l+\frac{1}{2}\right)^2 ,
  \label{eq:total_potential}
\end{equation}
where $\mu$ is the reduced mass and the last term is the Langer-modified centrifugal barrier. The nuclear part is a deformed Woods--Saxon potential,
\begin{equation}
  V_N(r,\theta)=
  -\frac{V_0}{1+\exp\{[r-R_N(\theta)]/a_N(\theta)\}},
  \label{eq:woods_saxon}
\end{equation}
with
\begin{equation}
  R_N(\theta)=R_0\left[1+\beta_2Y_{20}(\theta)+\beta_4Y_{40}(\theta)\right].
  \label{eq:surface}
\end{equation}
The angular dependence of the diffuseness\cite{PhysRevC.88.064327} is written as
\begin{equation}
\begin{split}
  a_N(\theta)=&\,a_0
  \left[1+\left(\frac{1}{R_N}\frac{\partial R_N}{\partial\theta}\right)^2\right]^{1/2}\\
  &\times
  \left[1+\tilde{\beta}_2Y_{20}(\theta)+\tilde{\beta}_4Y_{40}(\theta)\right],
\end{split}
  \label{eq:diffuseness}
\end{equation}
and the present calculations set $\tilde{\beta}_2=\tilde{\beta}_4=0$ unless otherwise stated. The spin-orbit term\cite{PhysRevC.60.054318} is
\begin{equation}
  V_{so}(r)=V_{so0}\lambda_\pi^2
  \frac{1}{r}\frac{\dif}{\dif r}
  \left[1+\exp\left(\frac{r-R_{so}}{a_{so}}\right)\right]^{-1}
  \vec{\sigma}\cdot\vec{l},
  \label{eq:spinorbit}
\end{equation}
where $\lambda_\pi^2=2~\fm^2$ and
\begin{equation}
  \vec{\sigma}\cdot\vec{l}
  = j(j+1)-l(l+1)-\frac{3}{4}.
\end{equation}
In addition, $R_{so}$ and $a_{so}$ take the same form as that of $R_N(\theta)$ and $a_{N}(\theta)$.

The Coulomb field is evaluated with a finite charge radius $R_{C0}$ under the assumption that the daughter nucleus is spherical and has a uniformly distributed charge
\begin{equation}
  V_C(r)=
  \begin{cases}
  \displaystyle \frac{Z_D Z_pe^2}{2R_{C0}}
  \left(3-\frac{r^2}{R_{C0}^2}\right), & r\le R_{C0},\\[6pt]
  \displaystyle \frac{Z_DZ_pe^2}{r}, & r>R_{C0}.
  \end{cases}
  \label{eq:coulomb}
\end{equation}
However, the deformed Coulomb potential\cite{bykhalo2021descriptionshapemediumheavy} should be obtained through multipole expansion instead of changing the radius directly
\begin{equation}
\begin{aligned}
& V_C(r,\theta,\phi)=\frac{3Z_DZ_pe^2}{R_{c0}^3}\sum_{\lambda=0}^{\infty}(2\lambda+1)^{-1}\sum_{\mu=-\lambda}^{\lambda}Y_{\lambda\mu}(\theta,\phi)\\
& \times\int_0^{2\pi}d\phi'\int_0^{\pi}Y_{\lambda\mu}^*(\theta',\phi')K_{\lambda}(r,\theta',\phi')\sin\theta'd\theta',
\end{aligned}
\end{equation}
where the radial kernel is
\begin{equation}
  K_{\lambda}(r,\theta,\phi)=
  \begin{cases}
  \dfrac{(2\lambda+1)r^2}{(\lambda+3)(\lambda-2)}-\dfrac{r^{\lambda}(\lambda-2)^{-1}}{R_c^{\lambda-2}}, & r\le R_c,\ \lambda\ne 2,\\[4pt]
  \dfrac{r^2}{5}+r^2\ln(\dfrac{R_c}{r}), & r\le R_c,\ \lambda=2,\\[4pt]
  \dfrac{1}{\lambda+3}\dfrac{R_c^{\lambda+3}}{r^{\lambda+1}}, & r>R_c,
  \end{cases}
\end{equation}
with the deformed charge radius $R_c(\theta,\phi)=R_{c0}(1+\beta_2Y_{20}(\theta)+\beta_4Y_{40}(\theta))=R_c(\theta)$ (no $\phi$ dependence since $m=0$), and considering only $\mu=0$ terms due to axial symmetry, the final form of the deformed Coulomb potential is
\begin{equation}
\begin{aligned}
V_C(r,\theta)=&\frac{3Z_DZ_pe^2}{R_{c0}^3}\sum_{\lambda=0}^{\infty}\frac{2\pi}{2\lambda+1}Y_{\lambda 0}(\theta) \\
&\times 2\int_0^{\pi/2}Y_{\lambda 0}^*(\theta')K_{\lambda}(r,\theta')\sin\theta'd\theta'.
\end{aligned}
\end{equation}
When restricting to $\lambda=0,2,4$, the expansion involves three terms.

Since the deformed potential depends explicitly on the angle $\theta$, the observable proton decay width consequently becomes a function of this deformation angle.

\subsection{The WKB method}
In the semi-classical WKB framework, the decaying proton is treated as a single particle moving within a deformed potential core. Due to the shape deformation, the interaction potential $V(r,\theta)$ becomes orientation-dependent. 

For a given orientation $\theta$, the classical turning points $r_1(\theta)$ (the inner turning point), $r_2(\theta)$ (the outer turning point of the well), and $r_3(\theta)$ (the outer barrier turning point) are determined as numerical solutions of the equation 
\begin{equation}
  Q_p = V(r,\theta),
\end{equation}
where $Q_p$ denotes the proton-decay energy. Consequently, the orientation-dependent single-particle width $\Gamma_{\mathrm{sp}}(\theta)$ can be directly expressed as
\begin{equation}
  \Gamma_{\mathrm{sp}}(\theta) = F(\theta)\frac{\hbar^2}{4\mu} \exp\left[-2\int_{r_2(\theta)}^{r_3(\theta)} k(r,\theta)\,\dif r\right],
  \label{eq:sp_width}
\end{equation}
where $\mu$ is the reduced mass of the proton-daughter system, and the normalization factor $F(\theta)$ inside the potential well is approximated in the simplified form below
\begin{equation}
  F(\theta) = \left[\int_{r_1(\theta)}^{r_2(\theta)} \frac{\dif r}{2k(r,\theta)}\right]^{-1}.
\end{equation}
In the equations above, the local wave number $k(r,\theta)$ within each region is given by
\begin{equation}
  k(r,\theta) = \sqrt{\frac{2\mu}{\hbar^2}\left|Q_p-V(r,\theta)\right|}.
\end{equation}

Assuming axial and reflection symmetry for the deformed system, the total single-particle width $\Gamma_{\mathrm{sp}}$ is obtained by performing an angular average over the orientation space
\begin{equation}
  \Gamma_{\mathrm{sp}} = \int_0^{\pi/2}\Gamma_{\mathrm{sp}}(\theta)\sin\theta\,\dif\theta .
  \label{eq:angle_average}
\end{equation}
Finally, the experimental half-life $T_{1/2}$ is connected to the calculated total single-particle width through the spectroscopic factor $P_0$\cite{PhysRevC.56.1762,zhang2024analyticformulaprotonradioactivity}, which phenomenologically encapsulates the nuclear structure information,
\begin{equation}
  T_{1/2} = \frac{\hbar\ln 2}{P_0\Gamma_{\mathrm{sp}}}.
  \label{eq:half_life}
\end{equation}

\subsection{Sensitivity measures}

To systematically quantify which model parameters dominate the calculated half-lives, a sensitivity analysis is performed across the investigated systems. Rather than relying on logarithmic scaling, the signed relative sensitivity $S_i^{\mathrm{signed}}$ for a specific parameter $\alpha_i$ is explicitly defined as a percentage variation based on the posterior distribution:
\begin{equation}
  S_i^{\mathrm{signed}} \approx \frac{1}{2}\frac{T_{1/2}(q_{84,i})-T_{1/2}(q_{16,i})}{T_{1/2}(\mathrm{median})}\times 100\%,
  \label{eq:sensitivity}
\end{equation}
where $T_{1/2}(q_{16,i})$ and $T_{1/2}(q_{84,i})$ represent the calculated half-lives evaluated by shifting the parameter $\alpha_i$ to its 16th and 84th percentiles of the posterior distribution, respectively, while all other parameters are held fixed at their median reference values. The values $q_{16,i}$ and $q_{84,i}$ denote the 16th and 84th percentiles (quantiles) of the marginal posterior distribution for the specific parameter $\alpha_i$, respectively. These two quantiles correspond to the boundaries of a $68\%$ Bayesian credible interval (the Bayesian analogue of a traditional $\pm 1\sigma$ standard deviation range), ensuring that the parameter perturbations $\Delta \alpha_i$ are naturally normalized relative to their statistically permissible ranges. The denominator $T_{1/2}(\mathrm{median})$ represents the baseline half-life calculated by setting all model parameters simultaneously to their respective posterior median values.

\section{Calculation and Discussion}
\label{sec:calc}
On the basis of the theoretical framework established above, the present section applies the deformed proton-decay model to a selected set of proton emitters in the region between the proton magic numbers $Z=50$ and $Z=82$. In the current analysis, attention is chiefly focused on odd-$A$ nuclei, for which the proton-emission process can be treated within the single-particle picture introduced in Sec.~\ref{sec:theory}. The relevant model parameters entering the deformed potential are calibrated in a Bayesian manner. Specifically, Markov-chain Monte Carlo sampling is performed with the Python package \texttt{emcee}\cite{goodman2010ensemble,Foreman-Mackey_2013}, allowing the posterior distributions of the deformation-model parameters to be explored and propagated to the calculated proton-decay half-lives. The resulting parameter posteriors, half-life systematics, and sensitivity measures are discussed below.

\subsection{Posterior distributions and parameter correlations}
The Bayesian calibration provides not only optimal parameter estimates, but also the full posterior probability distributions and the correlations among the model parameters. To illustrate the typical behavior of the present MCMC analysis, some representative odd-$A$ proton emitters, \nuc{109}{I}, \nuc{145}{Tm}, and \nuc{171}{Au}, are selected from the investigated region between the proton magic numbers $Z=50$ and $Z=82$. These nuclei cover different mass regions of the dataset and therefore provide useful examples for examining how the deformed-potential parameters are constrained by the proton-decay observables. 

The posterior medians and 68\% credible intervals for all model parameters across the investigated $Z=50$--$82$ region are compiled in Tables~\ref{tab:posterior_summary} and \ref{tab:posterior_spinorbit} of Appendix~\ref{app:posteriors}. In addition, the corresponding MCMC diagnostics for the remaining nuclei are displayed in Figs.~\ref{fig:app_diagnostics_113Cs}--\ref{fig:app_diagnostics_176Tl}.

\begin{figure}[htbp]
  \centering
  \includegraphics[width=\columnwidth]{{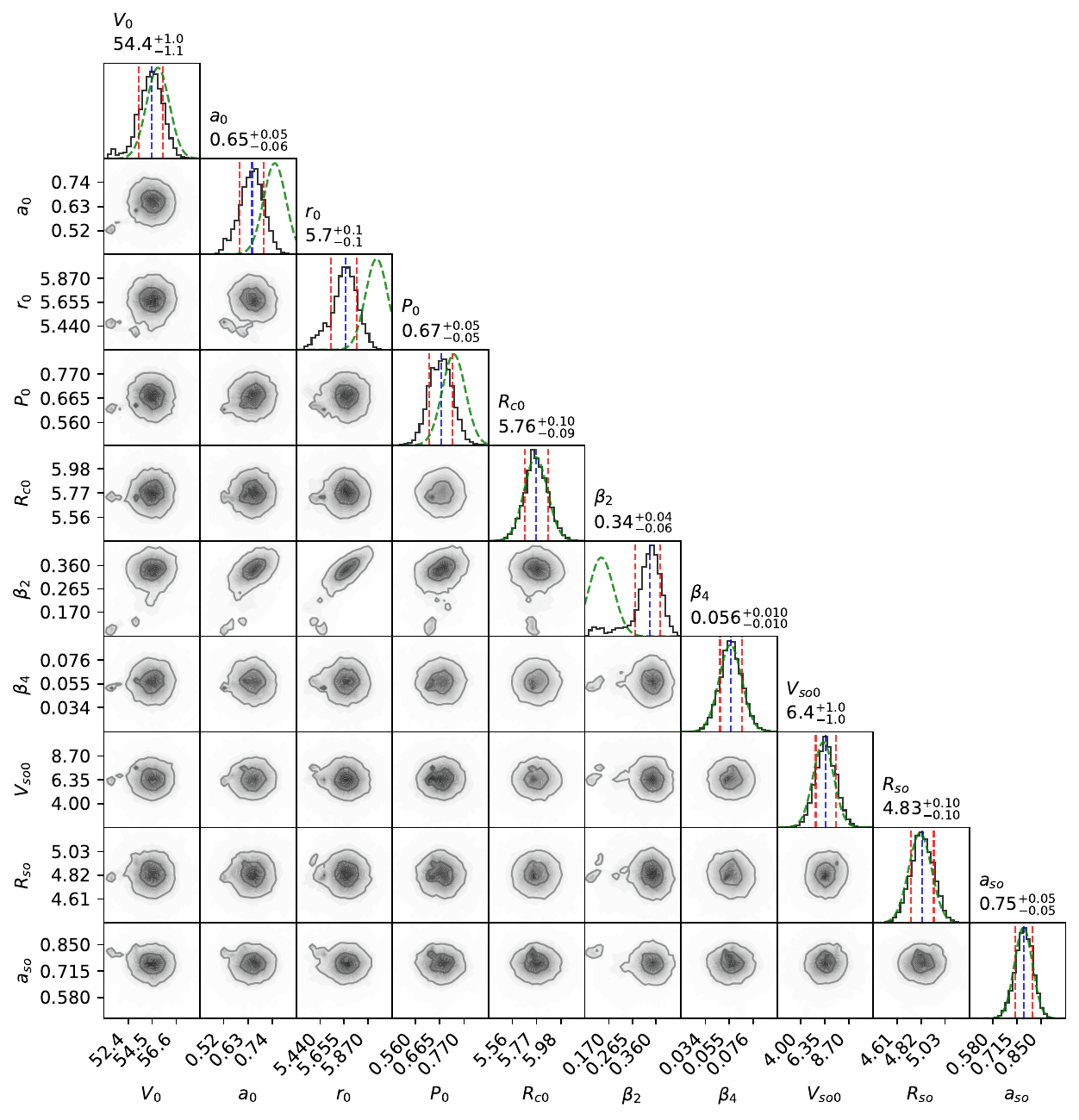}}
  \caption{Marginal and joint posterior distributions of the deformed-potential parameters obtained from MCMC sampling for \nuc{109}{I}. The diagonal panels show the one-dimensional marginal posteriors, while the off-diagonal panels display the corresponding two-dimensional joint probability distributions. In the one-dimensional marginal panels, the green dashed curves denote the prior distributions assigned to the corresponding parameters, and the red vertical lines mark the posterior quantiles associated with the $1\sigma$ credible interval.}
  \label{fig:posterior_109I}
\end{figure}

\begin{figure}[htbp]
  \centering
  \includegraphics[width=\columnwidth]{{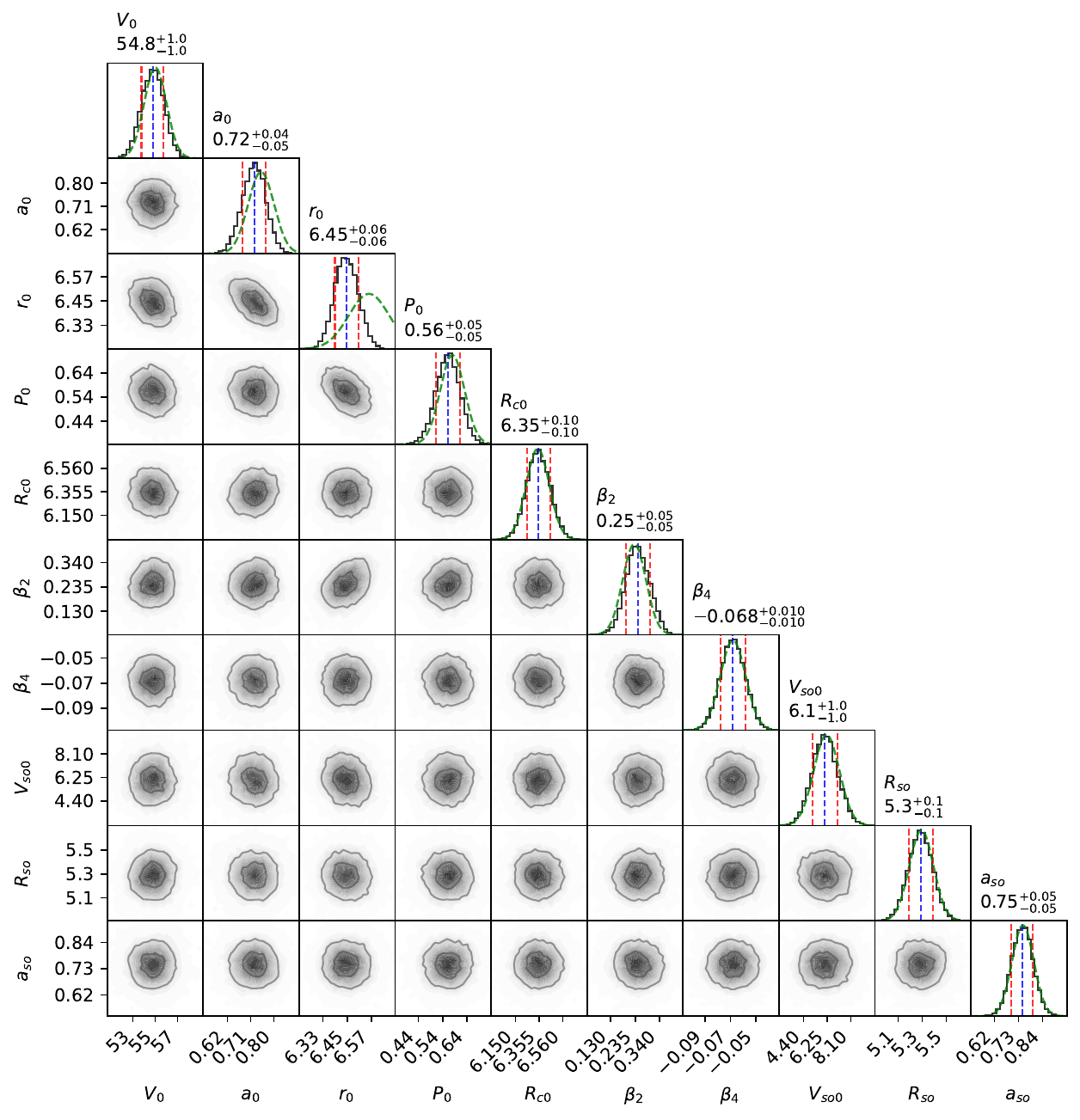}}
  \caption{Same as Fig.~\ref{fig:posterior_109I}, but for \nuc{145}{Tm}.}
  \label{fig:posterior_145Tm}
\end{figure}

\begin{figure}[htbp]
  \centering
  \includegraphics[width=\columnwidth]{{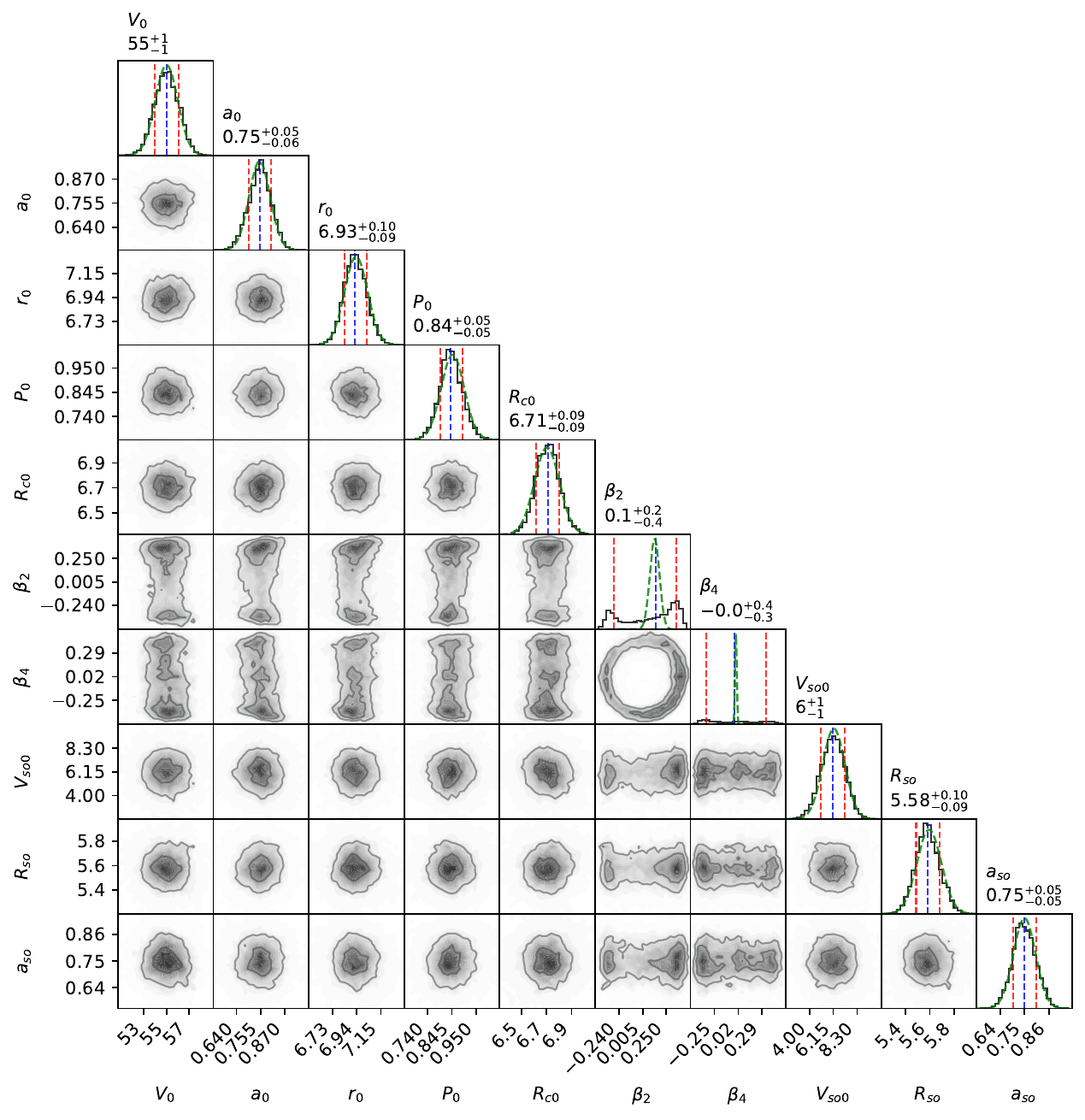}}
  \caption{Same as Fig.~\ref{fig:posterior_109I}, but for \nuc{171}{Au}.}
  \label{fig:posterior_171Au}
\end{figure}

Figures~\ref{fig:posterior_109I}--\ref{fig:posterior_171Au} present the posterior distributions sampled utilizing the affine-invariant ensemble sampler implemented in the \texttt{emcee} Python package. The diagonal panels show the marginalized posterior probability distributions for each individual parameter, while the off-diagonal panels display the corresponding pairwise joint posterior densities. Notably, the two-dimensional projections on the $\beta_2$--$\beta_4$ plane exhibit nearly circular contours, indicating that the quadrupole and hexadecapole deformations are statistically decoupled under current experimental constraints. In contrast, the joint distributions combining these shape parameters with the core potential parameters (such as $V_0$ and $r_0$) show rectangular or box-like profiles rather than tilted ellipses. This rectangular shape reflects an asymmetry in parameter sensitivity: while the core potential parameters are rigidly constrained within narrow intervals, the deformation parameters vary over a wider range without clear correlations, confirming that the model parameters are free of strong degeneracies.

\begin{figure}[htbp]
  \centering
  \includegraphics[width=\columnwidth]{{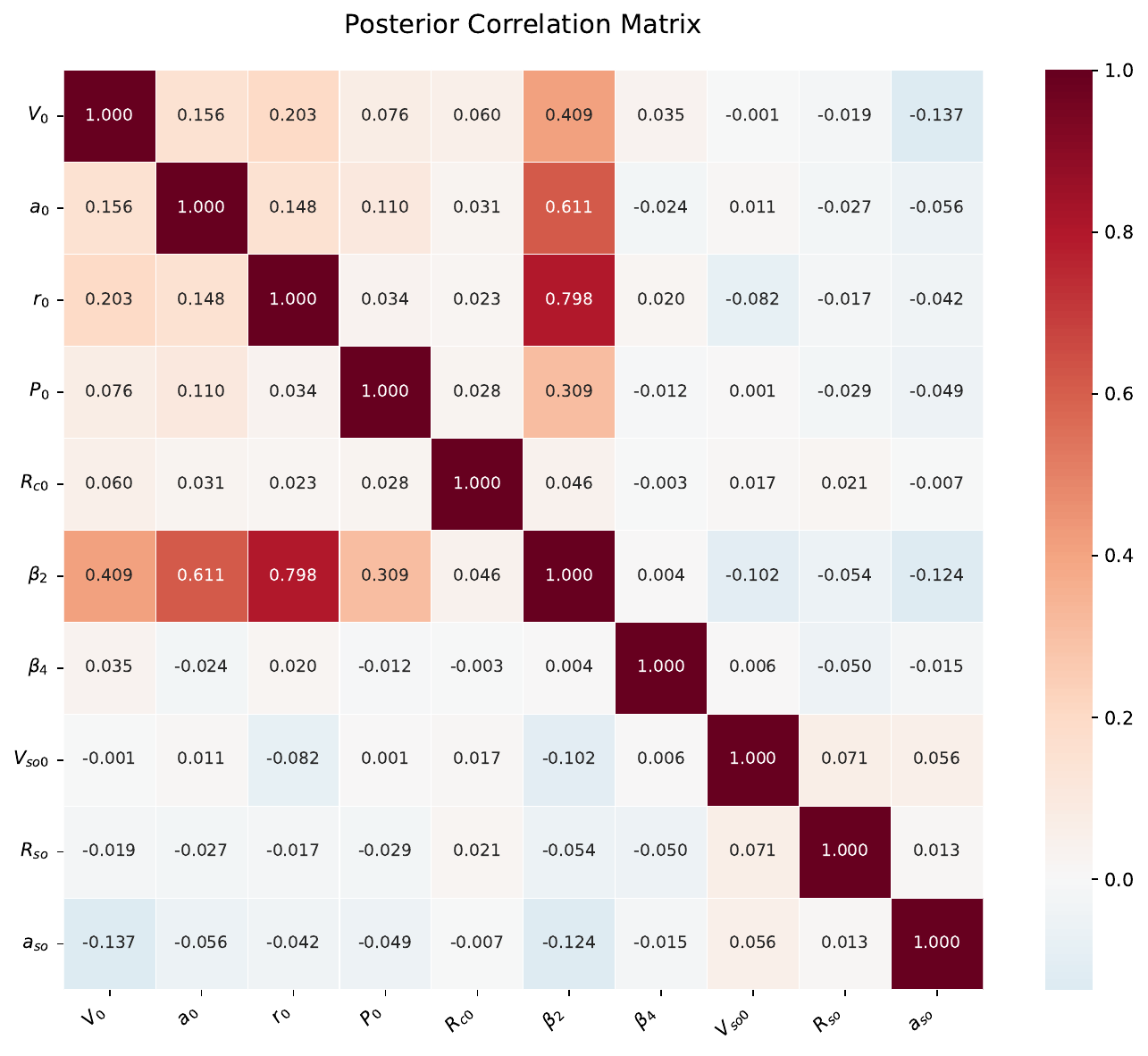}}
  \caption{Posterior Pearson correlation matrix for \nuc{109}{I}. The color scale quantifies the degree of linear correlation between pairs of sampled model parameters.}
  \label{fig:correlation_109I}
\end{figure}

\begin{figure}[htbp]
  \centering
  \includegraphics[width=\columnwidth]{{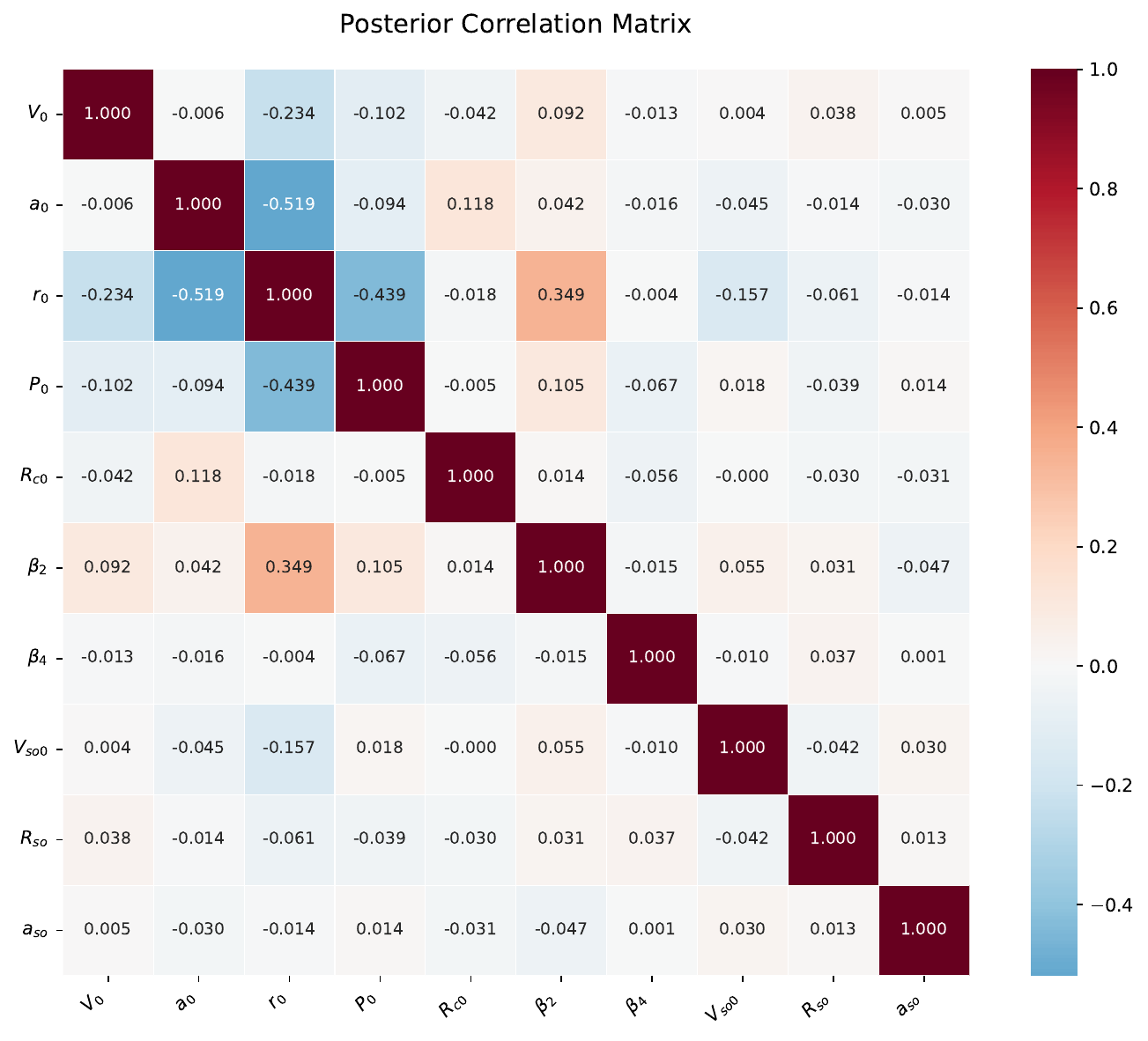}}
  \caption{Same as Fig.~\ref{fig:correlation_109I}, but for \nuc{145}{Tm}.}
  \label{fig:correlation_145Tm}
\end{figure}

\begin{figure}[htbp]
  \centering
  \includegraphics[width=\columnwidth]{{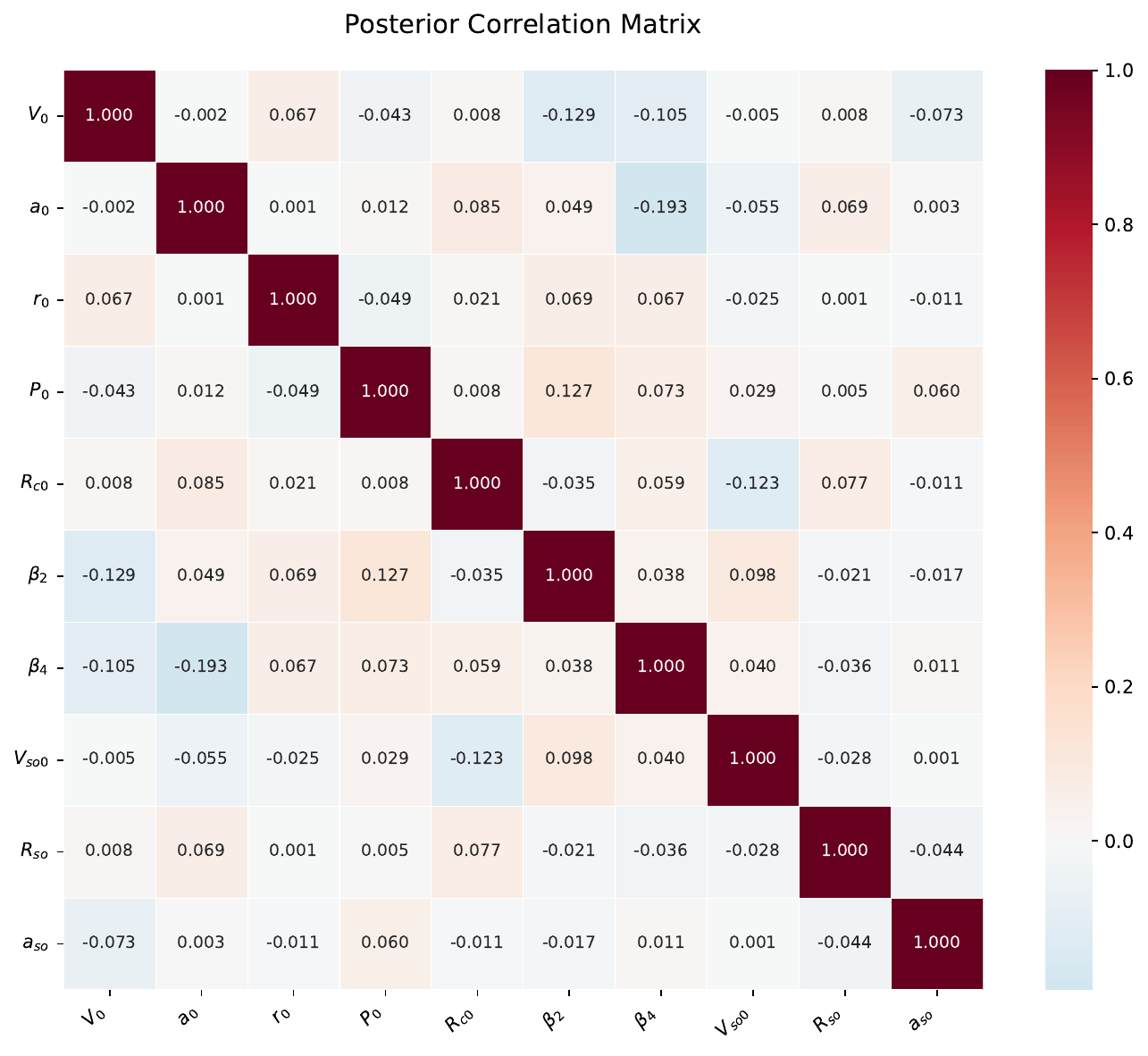}}
  \caption{Same as Fig.~\ref{fig:correlation_109I}, but for \nuc{171}{Au}.}
  \label{fig:correlation_171Au}
\end{figure}

The corresponding Pearson correlation matrices are presented in Figs.~\ref{fig:correlation_109I}--\ref{fig:correlation_171Au}. Across the investigated region, the off-diagonal matrix elements remain consistently small for almost all parameter pairs, reinforcing the absence of significant parameter degeneracies. This global lack of correlation is particularly evident in the $\beta_2$--$\beta_4$ sector, where the linear correlation coefficients drop close to zero, validating the statistical decoupling of multipole shape degrees of freedom. Furthermore, the near-zero correlations between these deformation parameters and the radial geometry parameters ($V_0, r_0$) underscore that the surface fluctuations and the bulk potential configuration operate as independent physical mechanisms in dictating the proton emission barrier.

\subsection{Systematic evolution of potential parameters}

\begin{figure}[htbp]
  \centering
  \includegraphics[width=\columnwidth]{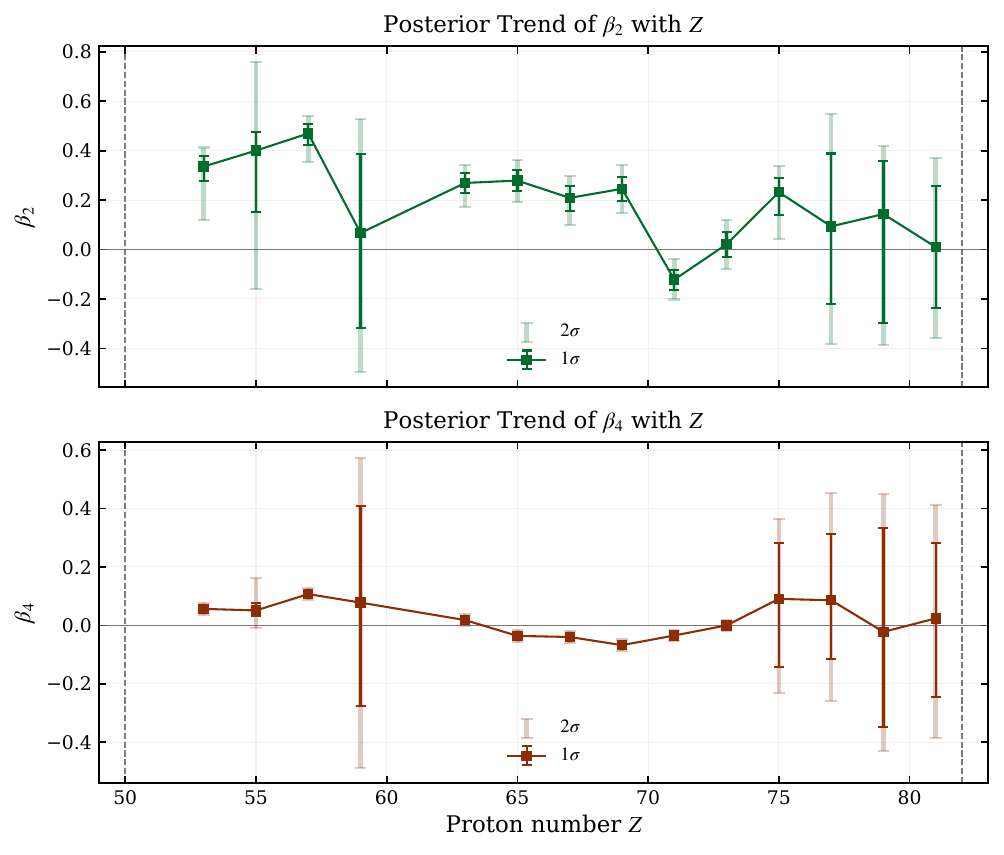}
  \caption{Systematic evolution of the quadrupole deformation parameter $\beta_2$ and the hexadecapole deformation parameter $\beta_4$ as functions of proton number $Z$ for the proton emitters considered in this work.}
  \label{fig:deformation_parameters_z}
\end{figure}

Figure~\ref{fig:deformation_parameters_z} illustrates the systematic evolution and posterior uncertainty distributions of the quadrupole ($\beta_2$, upper panel) and hexadecapole ($\beta_4$, lower panel) deformation parameters across the $Z=53$--81 sequence. The $1\sigma$ and $2\sigma$ shaded bands quantify the localized parameter uncertainties, providing statistical insight into the underlying nuclear shell effects.

The quadrupole deformation exhibits a clear transition around $Z=71$ driven by shell effects. The lighter mass sector ($Z=53$--69) lies in the mid-shell region, where strong collective forces lead to stable prolate shapes and narrow uncertainty bands. In contrast, the heavier sector ($Z \ge 71$) approaches the $Z=82$ magic closure. Near this magic shell, the underlying potential softens, causing a noticeable widening of the posterior uncertainty bands. This shell-driven behavior triggers a highly dynamic transitional zone, where the nuclear geometry oscillates between oblate shapes (at $Z=71, 81$) and weakly deformed prolate structures ($Z=73$--79), making the experimental half-lives less sensitive to specific deformation values.

The hexadecapole parameter $\beta_4$ exhibits a separate evolutionary pathway modulated by shell dynamics. Driven by the Bayesian constraints, the posterior distributions restrict $\beta_4$ into distinct geometric regimes across the $Z$ sequence. Specifically, in the lighter mass sector ($Z < 65$) and the heavier sector near the shell boundary ($Z > 73$), the $\beta_4$ values are well-constrained within the positive domain ($\beta_4 > 0$). In the intermediate region ($Z = 65$--73), however, the posterior median predictions shift below zero ($\beta_4 < 0$). Furthermore, similar to the behavior observed for $\beta_2$, a clear shell effect manifests in the uncertainty propagation: the $\sigma$ intervals of $\beta_4$ systematically expand as the nuclei approach the $Z=82$ magic closure, reflecting a softening of the underlying potential compared to the tighter constraints found in regions far from the magic shell.

\subsection{Systematic evolution of global parameter sensitivities}

To systematically assess the relative importance of each model parameter in governing the predicted half-lives across the investigated nuclear region, a global sensitivity analysis was performed by computing the absolute values of the signed relative sensitivities $|S_i|$ for all parameters in each nucleus, as defined in Eq.~(\ref{eq:sensitivity}). This analysis quantifies the fractional change in the calculated half-life that results from a $1\sigma$ variation in each parameter within its Bayesian credible interval, while holding all other parameters at their posterior median values.

\begin{figure}[htbp]
  \centering
  \includegraphics[width=\columnwidth]{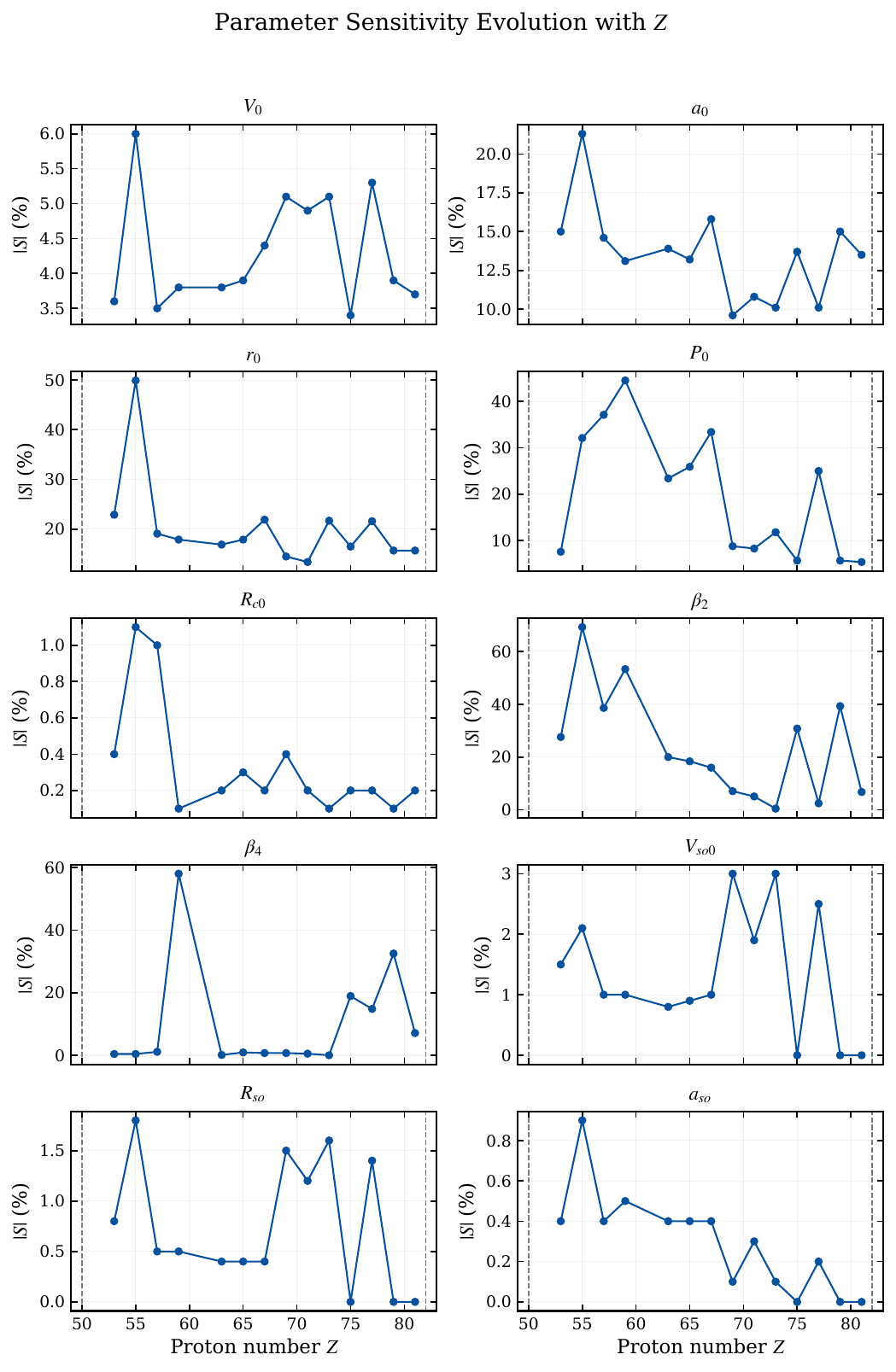}
  \caption{Evolution of parameter sensitivities across the investigated proton emitters as a function of proton number $Z$. Each panel displays the absolute value of the signed relative sensitivity $|S_i|$ for the model parameters, evaluated at the posterior median for each nucleus in the $Z=53$--81 sequence. Nuclear radius ($r_0$), spectroscopic factor ($P_0$), and multipole deformation parameters ($\beta_2$, $\beta_4$) emerge as the primary drivers of half-life uncertainty, whereas spin-orbit and Coulomb radius parameters exhibit highly suppressed or localized influences.}
  \label{fig:sensitivity_evolution_z}
\end{figure}

The global sensitivity hierarchy reveals a dynamic, mass-dependent interplay between nuclear bulk properties and multipole shape degrees of freedom. Parameters governing the nuclear geometry—specifically the nuclear radius $r_0$ and the quadrupole deformation $\beta_2$—act as the primary drivers of half-life uncertainty across most of the sequence, albeit with shifting relative dominance. The radial parameter $r_0$ exhibits an exceptional sensitivity peak near $Z=55$ ($|S_{r_0}| \approx 50\%$), driven by the exponential dependence of the tunneling probability on the barrier width. Beyond this peak, $|S_{r_0}|$ stabilizes into a persistent baseline of approximately $15\%$--$20\%$ across the remainder of the $Z$ sequence. 

The quadrupole deformation $\beta_2$ displays a pronounced structural evolution. In the lighter mass sector ($Z \le 59$), its sensitivity escalates dramatically, peaking at $|S_{\beta_2}| \approx 70\%$ in \nuc{113}{Cs}. This amplified sensitivity directly reflects the robust collective deformation in this mid-shell domain, where the decay width is acutely sensitive to the anisotropic surface orientation. Following a systematic decline across the $Z=61$--73 interval, where $|S_{\beta_2}|$ drops below $10\%$, a prominent secondary resurgence occurs in the heavy transitional region, climbing back to $|S_{\beta_2}| \approx 30\%$--$40\%$ at $Z=75$ and $79$.

The diffuseness parameter $a_0$ ranks as a consistently important secondary contributor, maintaining a stable profile within $|S_{a_0}| \approx 10\%$--$22\%$ across the entire sequence. In contrast, the spectroscopic factor $P_0$ manifests intermediate-to-high sensitivity that is strongly tied to shell structure. It exhibits a major collective peak of $|S_{P_0}| \approx 45\%$ near $Z=59$ before undergoing a systematic quenching down to $\sim 5\%$ for $Z \ge 69$, punctuated only by a sharp localized recovery to $25\%$ at $Z=77$, which underscores the fluctuating single-particle orbital occupations near major shell closures.

The higher-order shape configurations and peripheral coupling potentials show tightly localized or heavily suppressed profiles. The Coulomb radius $R_{c0}$ remains strictly constrained to $|S_{R_{c0}}| < 1.1\%$, confirming that long-range electrostatic variations exert minimal perturbation on the tunneling rates compared to nuclear-force geometries. Similarly, the spin-orbit potential parameters ($V_{so0}, R_{so}, a_{so}$) are universally subordinate, staying well below $3\%$ throughout the sequence. 

Crucially, the hexadecapole deformation $\beta_4$ exhibits a unique, bimodal sensitivity landscape. Rather than being universally suppressed, $\beta_4$ becomes actively involved in two isolated structural windows: a sharp collective spike at $Z=59$ ($|S_{\beta_4}| \approx 58\%$ in \nuc{121}{Pr}) and a pronounced resurgence in the heavy regime ($Z=75$--79), peaking at $|S_{\beta_4}| \approx 33\%$ in \nuc{171}{Au}. In the intermediate mass interval ($Z=63$--73), the $\beta_4$ sensitivity is completely quenched to zero.

This systemic re-ordering of sensitivities exposes a profound structural transition in the potential energy landscape, highlighting a clean distinction between intrinsic parameter sensitivity and marginalized posterior uncertainties. In the intermediate-mass domain ($Z=63$--69), the rigid and well-deformed potential forms a highly localized, sharp multidimensional posterior hyper-volume devoid of parametric degeneracies. Although the global sensitivity of the hexadecapole parameter $\beta_4$ is extensively quenched to zero in this sector, the absolute lack of directional compensation pathways under strict, dominant radial and quadrupole constraints forces this higher-order deformation to converge rigidly within a remarkably narrow, well-defined posterior corridor, as observed in the structural landscape. 

Conversely, as the sequence approaches the $Z=82$ major shell closure ($Z \ge 75$), this sensitivity hierarchy flattens into a multi-dimensional transitional signature. In this heavy near-magic regime, the single-particle potential undergoes substantial softening, flattening the energy surface and amplifying the fractional impact of higher-order surface fluctuations on the proton tunneling probability. This physical amplification forces the decay width to become simultaneously and exceptionally sensitive to a concurrent combination of multipole configurations, manifested by the dramatic sensitivity resurgences of both $\beta_2$ and $\beta_4$. Crucially, because these active deformation modes can mutually compensate for one another within the flattened, shape-coexisting potential landscape, the enhanced sensitivity does not result in narrower parameter bounds; instead, this multidimensional parameter degeneracy geometrically inflates the marginalized posterior uncertainty bands at the terminal points of the sequence.

\subsection{Uncertainty quantification of proton-decay half-lives}

The posterior parameter distributions obtained from Bayesian calibration are propagated through the WKB framework to quantify predictive uncertainties in the calculated half-lives.

\begin{figure}[htbp]
  \centering
  \includegraphics[width=\columnwidth]{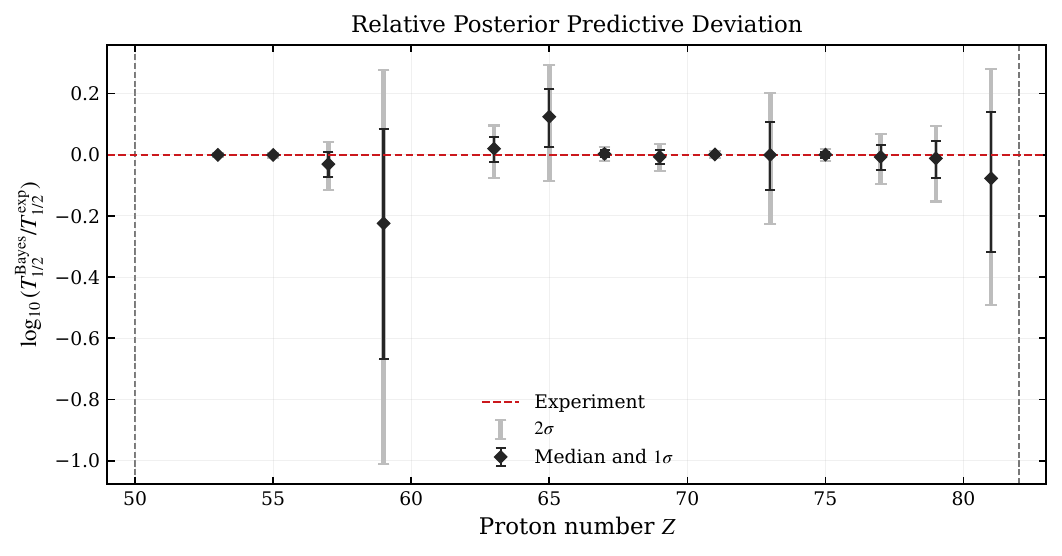}
  \caption{Relative posterior predictive deviation for proton-decay half-lives, expressed as $\log_{10}(T^{\text{Bayes}}_{1/2}/T^{\text{exp}}_{1/2})$. Black markers with error bars represent posterior median and $1\sigma$ credible interval; gray bands denote $2\sigma$ interval. }
  \label{fig:half_life_uq}
\end{figure}

Figure~\ref{fig:half_life_uq} displays the relative posterior predictive deviation $\log_{10}(T^{\text{Bayes}}_{1/2}/T^{\text{exp}}_{1/2})$ across the $Z=50$--82 region. Zero deviation (red dashed line) indicates perfect agreement with experiment. The posterior median and $1\sigma$ credible interval (black markers and error bars) are contrasted with the $2\sigma$ interval (gray bands).

The relative posterior predictive deviations displayed in Fig.~\ref{fig:half_life_uq} do not exhibit a monotonic, shell-driven trend across the proton-number sequence; instead, the uncertainty bands manifest a highly localized, cell-by-cell volatility. The widths of both the $1\sigma$ credible intervals and $2\sigma$ error bands fluctuated conspicuously along the $Z$ sequence, characterized by extremely tight convergence in several isotopes (e.g., $Z=53, 55, 67, 71$) interspersed with pronounced localized expansions, most notably at $Z=59$ and toward the terminal limit at $Z=81$. 

This non-uniform statistical pattern elegantly reflects the varying degrees of intrinsic parametric sensitivity and localized structural anomalies within the specific nuclei. For instance, the drastic error inflation observed at $Z=59$ directly correlates with the global sensitivity analysis, where several inputs ($ P_0, \beta_4$) concurrently hit their maximum sensitivity peaks, thereby exponentially amplifying minor parameteric variances at the final half-life output level. 

Crucially, despite the non-uniform magnitudes of the error bars across the sequence, the experimental half-lives remain comfortably encompassed within the quantified $1\sigma$ and $2\sigma$ intervals without systematic bias. This variation in uncertainty demonstrates that the framework does not suffer from parameter overfitting. Instead, it proves that the Bayesian full propagation successfully provides a highly differentiated, case-by-case diagnosis of the epistemic uncertainties based on local shell dynamics. Ultimately, the tight agreement between the posterior median predictions and the experimental data across multiple orders of magnitude demonstrates that the deformed-potential framework reliably captures the essential physics governing proton radioactivity.

\section{Conclusion}
\label{sec:conclusion}

In this work, a rigorous Bayesian uncertainty quantification (UQ) framework for proton radioactivity has been established by embedding the orientation-dependent WKB approximation within a deformed Woods--Saxon potential. Through Markov-chain Monte Carlo sampling constrained by experimental half-lives, the joint posterior probability distributions and parameter covariances are successfully extracted. Crucially, these statistical metrics enable a reliable propagation of parametric uncertainties to the calculated observables, providing well-defined posterior predictive distributions for proton-decay half-lives. Furthermore, a systematic ranking of parameter importances is achieved via a signed relative sensitivity index. The global sensitivity hierarchy reveals that the nuclear radius and the quadrupole deformation consistently emerge as the most dominant parameters governing the proton-decay dynamics across the $Z=50$--$82$ region.

Furthermore, the framework provides deep insights into the correlations between macroscopic observables and underlying shell structures. Near the $Z=82$ major shell closure, the systematic widening of the half-life $\sigma$ uncertainty bands are physically interpreted as a direct consequence of the structural softening of the nuclear potential, which triggers enhanced parametric degeneracies and shape coexistence. This successful mapping between empirical error profiles and potential energy configurations underlines the strong potential of the Bayesian UQ methodology to extract reliable nuclear structure information directly from decay systematics.

Additionally, the Bayesian analysis highlights the occurrence of sloppiness in the posterior distributions, most notably in \nuc{113}{Cs}. In this case, the joint posterior distributions exhibit pronounced elongated and tilted structures. This strong parameter correlation indicates significant degeneracies and sloppy directions, where different combinations of potential parameters can compensate each other while producing comparable decay widths. Such findings underscore the value of full uncertainty quantification beyond point estimates, providing deeper insight into the complex interplay between model parameters and nuclear structure information.

To further extend the scope of this framework, incorporating more comprehensive, microscopic frameworks such as advanced coupled-channels calculations or triaxial degrees of freedom represent a compelling next step. Such developments will yield deeper insights into the underlying physical parameter hierarchies and further solidify the predictive power of Bayesian inference for exotic nuclei near the proton drip line.

\begin{acknowledgments}
This work was supported by the National Natural Science Foundation of China (Grant No.~12475132 and No.~12535009), by the National Key R\&D Program of China (Contract No.~2023YFA1606503), and by the Fundamental Research Funds for the Central Universities. During this work, I use AI-based tools(large language model assistants) to help check the syntax and polish the text. All physics content and results were verified by myself, who is responsible for this work.
\end{acknowledgments}

\appendix

\section{Supplementary Posterior Distributions and Parameter Summary}
\label{app:posteriors}

To provide a comprehensive overview of the Bayesian results, additional posterior distributions and a global parameter summary are presented in this Appendix. For brevity, only representative systems are discussed in the main text.

The posterior medians and 68\% credible intervals for all model parameters across the investigated nuclei are summarized in Tables~\ref{tab:posterior_summary} and \ref{tab:posterior_spinorbit}.

\begin{table*}[htbp]
\centering
\caption{Posterior medians and 68\% credible intervals of the deformed-potential parameters (Woods-Saxon and deformation). All values are obtained from the marginal posterior samples of the Bayesian MCMC chains.}
\label{tab:posterior_summary}
\begin{ruledtabular}
\resizebox{\textwidth}{!}{%
\begin{tabular}{lccccccc}
Nucleus & $V_0$ (MeV) & $a_0$ (fm) & $r_0$ (fm) & $P_0$ & $R_{c0}$ (fm) & $\beta_2$ & $\beta_4$ \\
\hline
\nuc{109}{I}   & $54.42^{+1.00}_{-1.10}$ & $0.6481^{+0.0510}_{-0.0560}$ & $5.666^{+0.100}_{-0.130}$ & $0.6708^{+0.0500}_{-0.0520}$ & $5.764^{+0.100}_{-0.094}$ & $0.3357^{+0.0410}_{-0.0580}$ & $0.05649^{+0.01000}_{-0.00960}$ \\
\nuc{113}{Cs}  & $54.89^{+1.90}_{-1.50}$ & $0.6430^{+0.0750}_{-0.0750}$ & $5.665^{+0.190}_{-0.260}$ & $0.2549^{+0.0800}_{-0.0730}$ & $5.809^{+0.150}_{-0.210}$ & $0.4002^{+0.0760}_{-0.2500}$ & $0.05114^{+0.02700}_{-0.01700}$ \\
\nuc{117}{La}  & $54.69^{+0.99}_{-1.00}$ & $0.6894^{+0.0490}_{-0.0500}$ & $5.928^{+0.093}_{-0.095}$ & $0.1990^{+0.0630}_{-0.0670}$ & $5.905^{+0.100}_{-0.100}$ & $0.4696^{+0.0380}_{-0.0470}$ & $0.1072^{+0.0099}_{-0.0100}$ \\
\nuc{121}{Pr}  & $55.01^{+1.00}_{-1.00}$ & $0.7475^{+0.0460}_{-0.0480}$ & $6.161^{+0.100}_{-0.099}$ & $0.1260^{+0.0490}_{-0.0480}$ & $5.964^{+0.100}_{-0.097}$ & $0.06683^{+0.32000}_{-0.38000}$ & $0.07799^{+0.33000}_{-0.35000}$ \\
\nuc{131}{Eu}  & $55.18^{+0.97}_{-1.00}$ & $0.7840^{+0.0500}_{-0.0480}$ & $6.421^{+0.097}_{-0.089}$ & $0.1266^{+0.0300}_{-0.0280}$ & $6.129^{+0.099}_{-0.097}$ & $0.2694^{+0.0390}_{-0.0410}$ & $0.01811^{+0.00970}_{-0.00960}$ \\
\nuc{135}{Tb}  & $55.17^{+1.00}_{-1.00}$ & $0.7799^{+0.0490}_{-0.0490}$ & $6.479^{+0.094}_{-0.093}$ & $0.1208^{+0.0320}_{-0.0280}$ & $6.189^{+0.100}_{-0.100}$ & $0.2789^{+0.0420}_{-0.0440}$ & $-0.03603^{+0.00970}_{-0.00990}$ \\
\nuc{141}{Ho}  & $55.27^{+1.20}_{-1.10}$ & $0.7847^{+0.0680}_{-0.0510}$ & $6.595^{+0.140}_{-0.099}$ & $0.1039^{+0.0280}_{-0.0330}$ & $6.279^{+0.100}_{-0.100}$ & $0.2089^{+0.0470}_{-0.0530}$ & $-0.03973^{+0.00980}_{-0.01000}$ \\
\nuc{145}{Tm}  & $54.71^{+0.99}_{-0.94}$ & $0.7262^{+0.0460}_{-0.0470}$ & $6.444^{+0.059}_{-0.056}$ & $0.5587^{+0.0480}_{-0.0460}$ & $6.348^{+0.100}_{-0.110}$ & $0.2457^{+0.0490}_{-0.0500}$ & $-0.06774^{+0.00990}_{-0.00990}$ \\
\nuc{151}{Lu}  & $55.66^{+0.98}_{-0.94}$ & $0.8261^{+0.0460}_{-0.0470}$ & $6.962^{+0.058}_{-0.056}$ & $0.5495^{+0.0450}_{-0.0450}$ & $6.433^{+0.100}_{-0.099}$ & $-0.1221^{+0.0400}_{-0.0410}$ & $-0.03462^{+0.01000}_{-0.01000}$ \\
\nuc{155}{Ta}  & $55.03^{+0.98}_{-0.96}$ & $0.7493^{+0.0510}_{-0.0490}$ & $6.711^{+0.090}_{-0.087}$ & $0.4250^{+0.0480}_{-0.0500}$ & $6.479^{+0.097}_{-0.097}$ & $0.02133^{+0.05000}_{-0.05100}$ & $5.347e-05^{+0.00980000}_{-0.00980000}$ \\
\nuc{161}{Re}  & $54.97^{+0.90}_{-1.00}$ & $0.7470^{+0.0490}_{-0.0490}$ & $6.794^{+0.097}_{-0.099}$ & $0.8941^{+0.0500}_{-0.0520}$ & $6.558^{+0.100}_{-0.099}$ & $0.2321^{+0.0580}_{-0.0930}$ & $0.09072^{+0.19000}_{-0.23000}$ \\
\ensuremath{{}^{165\mathrm{m}}\mathrm{Ir}}  & $55.02^{+0.99}_{-1.10}$ & $0.7411^{+0.0470}_{-0.0480}$ & $6.824^{+0.090}_{-0.087}$ & $0.1741^{+0.0440}_{-0.0400}$ & $6.630^{+0.100}_{-0.100}$ & $0.09352^{+0.29000}_{-0.31000}$ & $0.08600^{+0.23000}_{-0.20000}$ \\
\nuc{171}{Au}  & $55.01^{+1.10}_{-1.10}$ & $0.7532^{+0.0520}_{-0.0560}$ & $6.927^{+0.099}_{-0.093}$ & $0.8399^{+0.0510}_{-0.0450}$ & $6.711^{+0.092}_{-0.094}$ & $0.1435^{+0.2100}_{-0.4400}$ & $-0.02221^{+0.36000}_{-0.32000}$ \\
\nuc{176}{Tl}  & $55.04^{+1.00}_{-1.00}$ & $0.7522^{+0.0500}_{-0.0500}$ & $7.009^{+0.096}_{-0.095}$ & $0.9274^{+0.0490}_{-0.0510}$ & $6.765^{+0.100}_{-0.100}$ & $0.009717^{+0.250000}_{-0.250000}$ & $0.02391^{+0.26000}_{-0.27000}$ \\
\end{tabular}%
1}
\end{ruledtabular}
\end{table*}

\begin{table*}[htbp]
\centering
\caption{Posterior medians and 68\% credible intervals of the spin-orbit potential parameters. All values are obtained from the marginal posterior samples of the Bayesian MCMC chains.}
\label{tab:posterior_spinorbit}
\begin{ruledtabular}
\begin{tabular}{lccc}
Nucleus & $V_{\rm so0}$ (MeV) & $R_{\rm so}$ (fm) & $a_{\rm so}$ (fm) \\
\hline
\nuc{109}{I}   & $6.429^{+1.000}_{-0.970}$ & $4.827^{+0.100}_{-0.099}$ & $0.7542^{+0.0480}_{-0.0500}$ \\
\nuc{113}{Cs}  & $6.729^{+1.100}_{-1.600}$ & $4.989^{+0.220}_{-0.170}$ & $0.7165^{+0.0780}_{-0.0990}$ \\
\nuc{117}{La}  & $6.227^{+1.000}_{-1.000}$ & $4.928^{+0.099}_{-0.098}$ & $0.7501^{+0.0500}_{-0.0470}$ \\
\nuc{121}{Pr}  & $6.208^{+0.920}_{-0.980}$ & $4.982^{+0.100}_{-0.097}$ & $0.7519^{+0.0510}_{-0.0510}$ \\
\nuc{131}{Eu}  & $6.246^{+1.000}_{-1.000}$ & $5.113^{+0.100}_{-0.099}$ & $0.7456^{+0.0520}_{-0.0500}$ \\
\nuc{135}{Tb}  & $6.241^{+0.970}_{-0.950}$ & $5.171^{+0.099}_{-0.097}$ & $0.7530^{+0.0490}_{-0.0500}$ \\
\nuc{141}{Ho}  & $6.231^{+1.000}_{-1.100}$ & $5.248^{+0.099}_{-0.099}$ & $0.7519^{+0.0490}_{-0.0540}$ \\
\nuc{145}{Tm}  & $6.063^{+0.960}_{-0.910}$ & $5.287^{+0.099}_{-0.099}$ & $0.7472^{+0.0520}_{-0.0510}$ \\
\nuc{151}{Lu}  & $6.467^{+0.980}_{-0.960}$ & $5.382^{+0.098}_{-0.098}$ & $0.7508^{+0.0500}_{-0.0480}$ \\
\nuc{155}{Ta}  & $6.196^{+1.000}_{-1.000}$ & $5.414^{+0.097}_{-0.099}$ & $0.7493^{+0.0510}_{-0.0480}$ \\
\nuc{161}{Re}  & $6.316^{+1.100}_{-1.000}$ & $5.478^{+0.100}_{-0.100}$ & $0.7490^{+0.0500}_{-0.0510}$ \\
\ensuremath{{}^{165\mathrm{m}}\mathrm{Ir}}  & $6.221^{+0.970}_{-0.970}$ & $5.520^{+0.100}_{-0.100}$ & $0.7424^{+0.0530}_{-0.0470}$ \\
\nuc{171}{Au}  & $6.131^{+1.100}_{-1.100}$ & $5.581^{+0.100}_{-0.092}$ & $0.7482^{+0.0550}_{-0.0490}$ \\
\nuc{176}{Tl}  & $6.192^{+1.000}_{-0.980}$ & $5.646^{+0.100}_{-0.099}$ & $0.7508^{+0.0510}_{-0.0500}$ \\
\end{tabular}%
\end{ruledtabular}
\end{table*}

\begin{figure*}[tbp]
  \centering
  \begin{minipage}[t]{0.49\textwidth}
    \centering
    \includegraphics[width=\linewidth]{{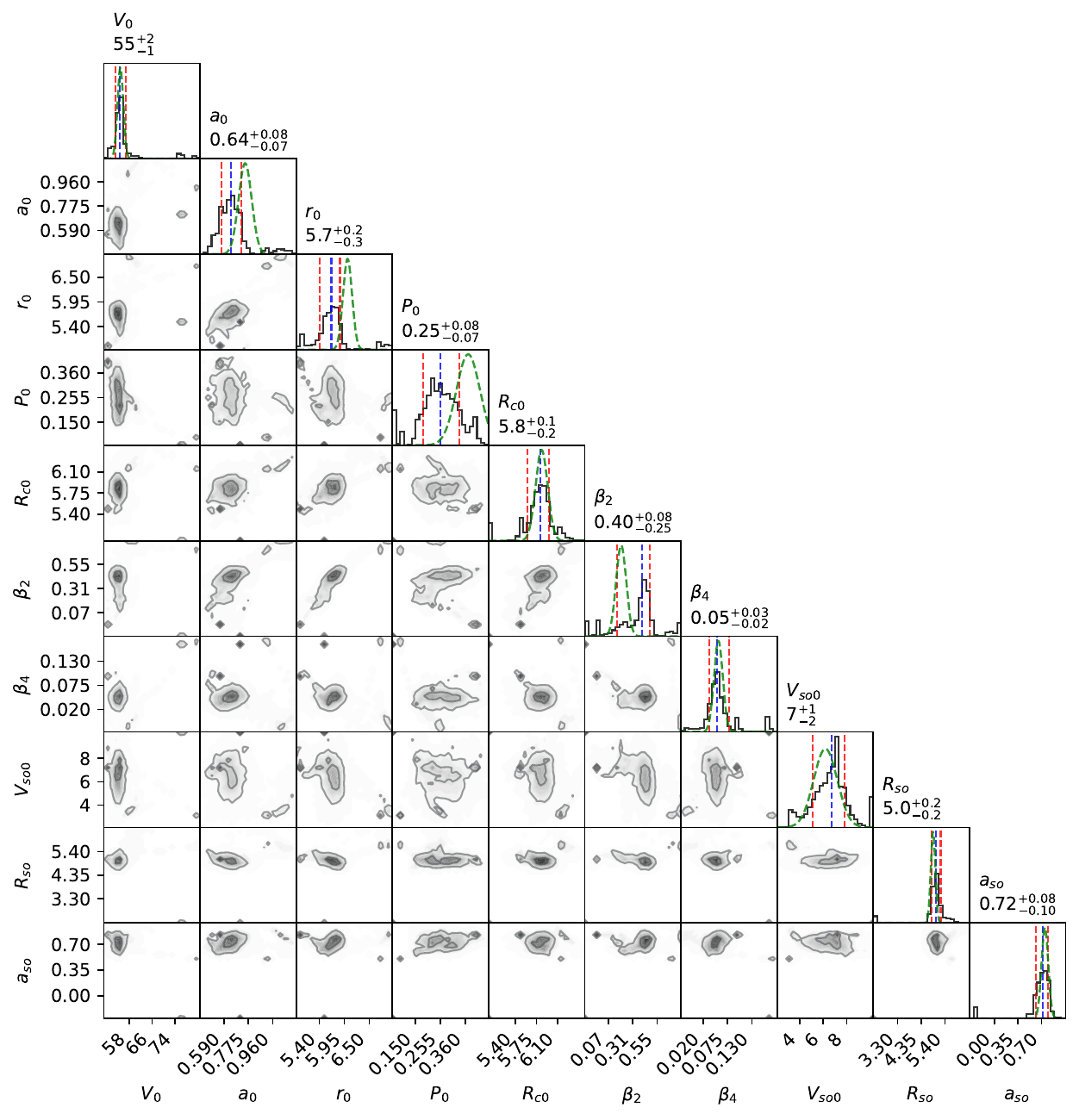}}
  \end{minipage}\hfill
  \begin{minipage}[t]{0.49\textwidth}
    \centering
    \includegraphics[width=\linewidth]{{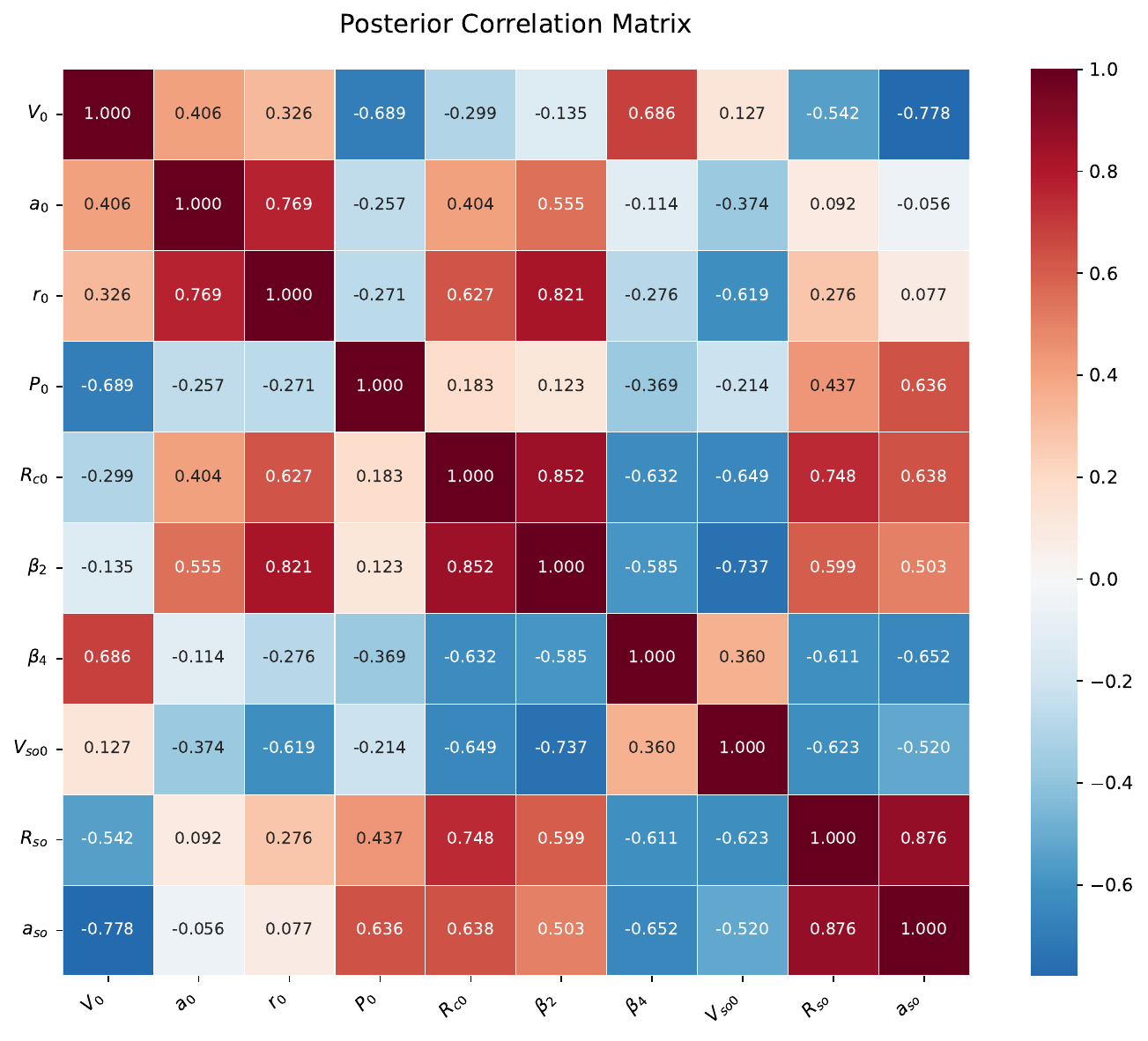}}
  \end{minipage}
  \caption{Supplementary posterior diagnostics for \nuc{113}{Cs}. Left: marginal and joint posterior distributions of the deformed-potential parameters obtained from MCMC sampling. Right: posterior Pearson correlation matrix of the sampled parameters, with the color scale indicating the linear correlation coefficient for each parameter pair.}
  \label{fig:app_diagnostics_113Cs}
  \label{fig:app_posterior_113Cs}
  \label{fig:app_correlation_113Cs}
\end{figure*}

\begin{figure*}[tbp]
  \centering
  \begin{minipage}[t]{0.49\textwidth}
    \centering
    \includegraphics[width=\linewidth]{{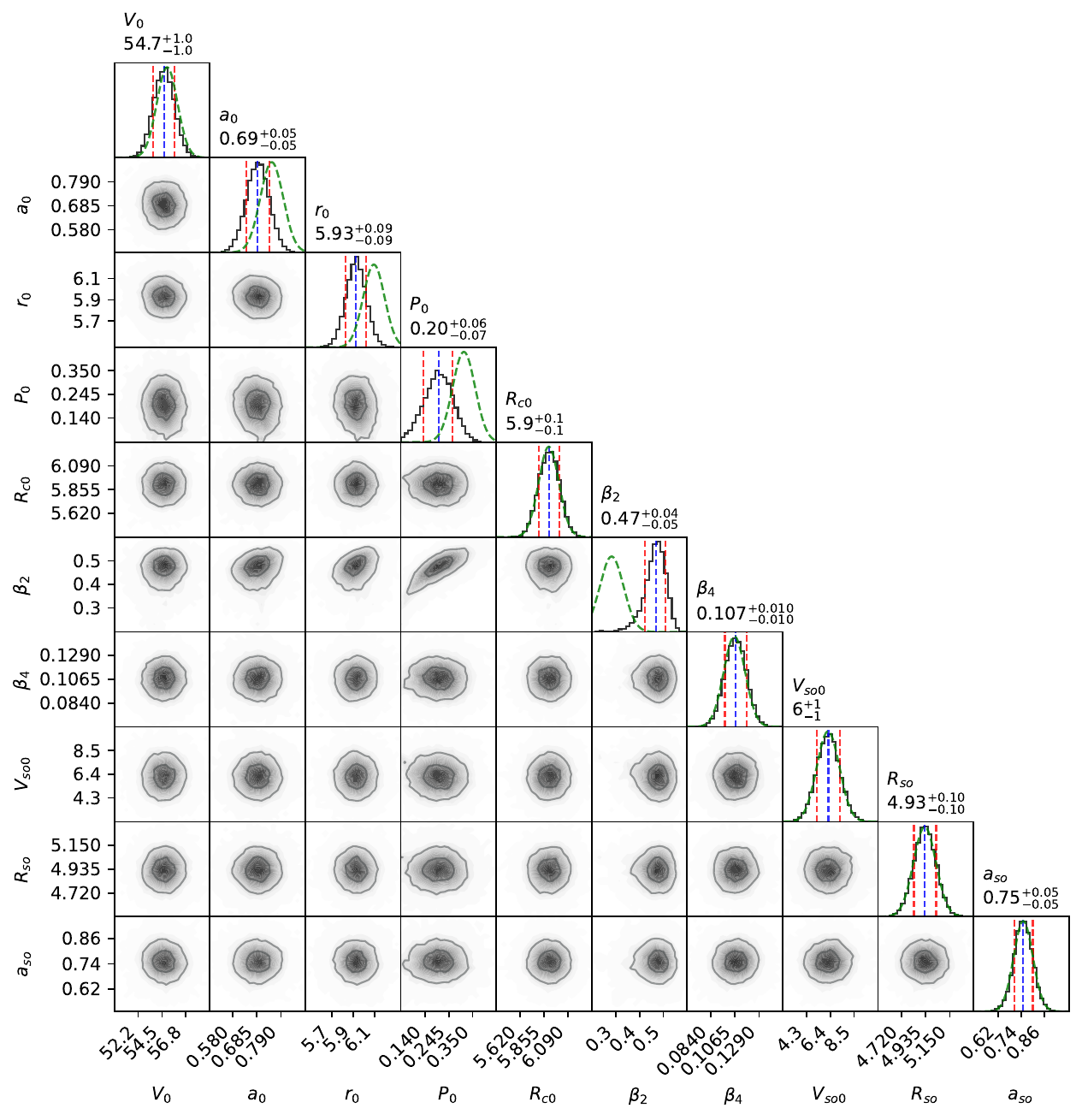}}
  \end{minipage}\hfill
  \begin{minipage}[t]{0.49\textwidth}
    \centering
    \includegraphics[width=\linewidth]{{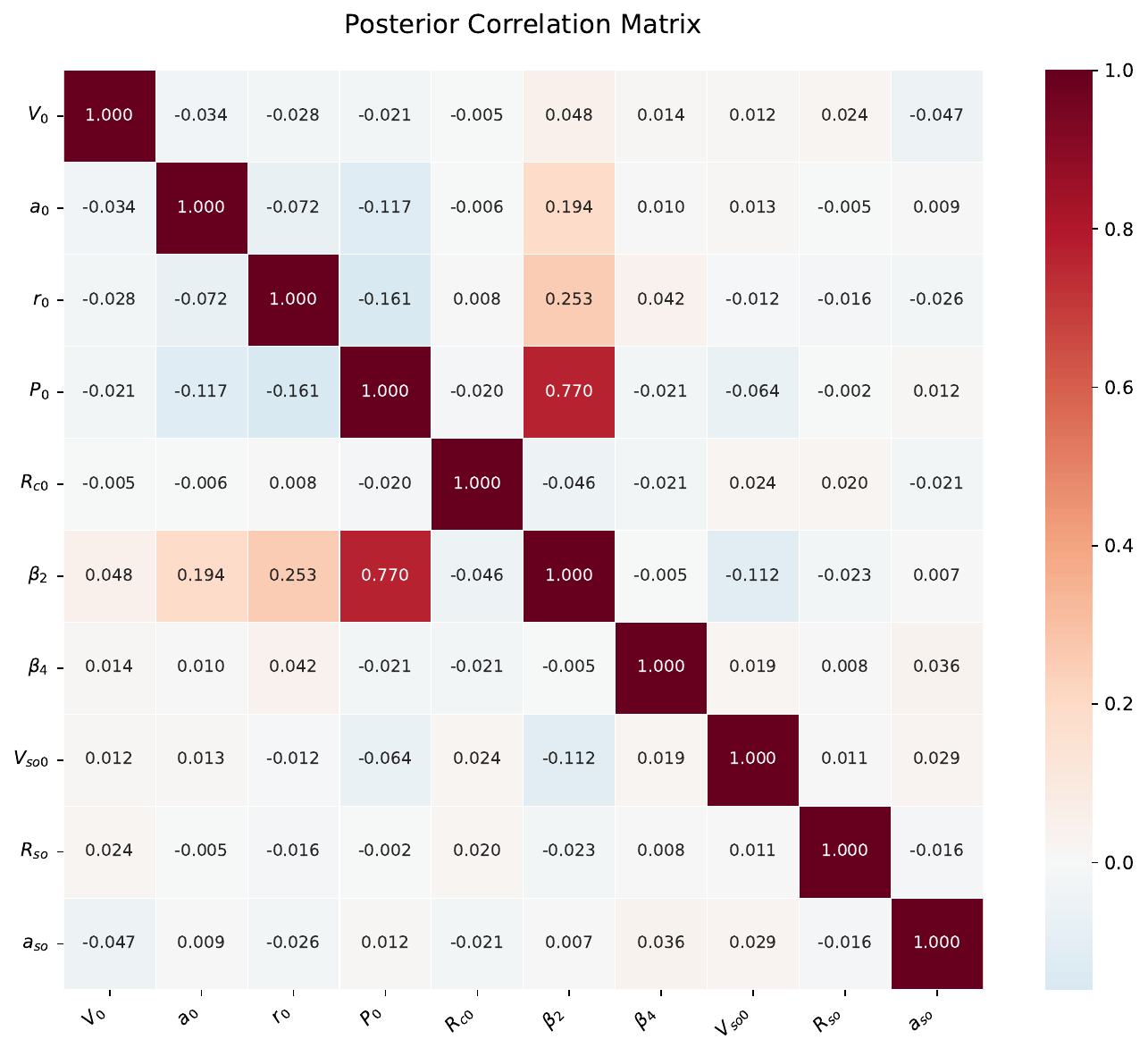}}
  \end{minipage}
  \caption{Supplementary posterior diagnostics for \nuc{117}{La}. Left: marginal and joint posterior distributions of the deformed-potential parameters obtained from MCMC sampling. Right: posterior Pearson correlation matrix of the sampled parameters, with the color scale indicating the linear correlation coefficient for each parameter pair.}
  \label{fig:app_diagnostics_117La}
  \label{fig:app_posterior_117La}
  \label{fig:app_correlation_117La}
\end{figure*}

\begin{figure*}[tbp]
  \centering
  \begin{minipage}[t]{0.49\textwidth}
    \centering
    \includegraphics[width=\linewidth]{{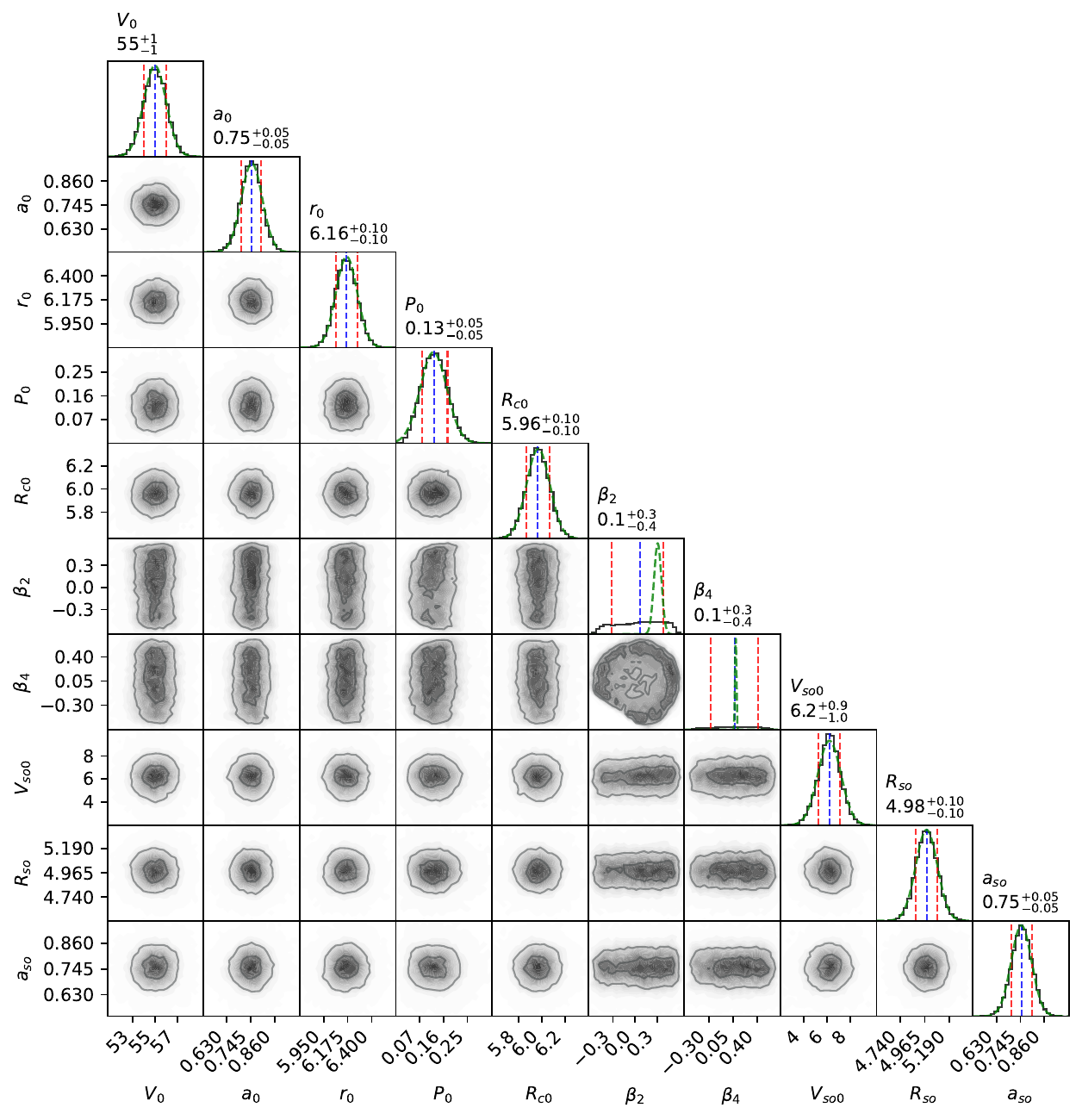}}
  \end{minipage}\hfill
  \begin{minipage}[t]{0.49\textwidth}
    \centering
    \includegraphics[width=\linewidth]{{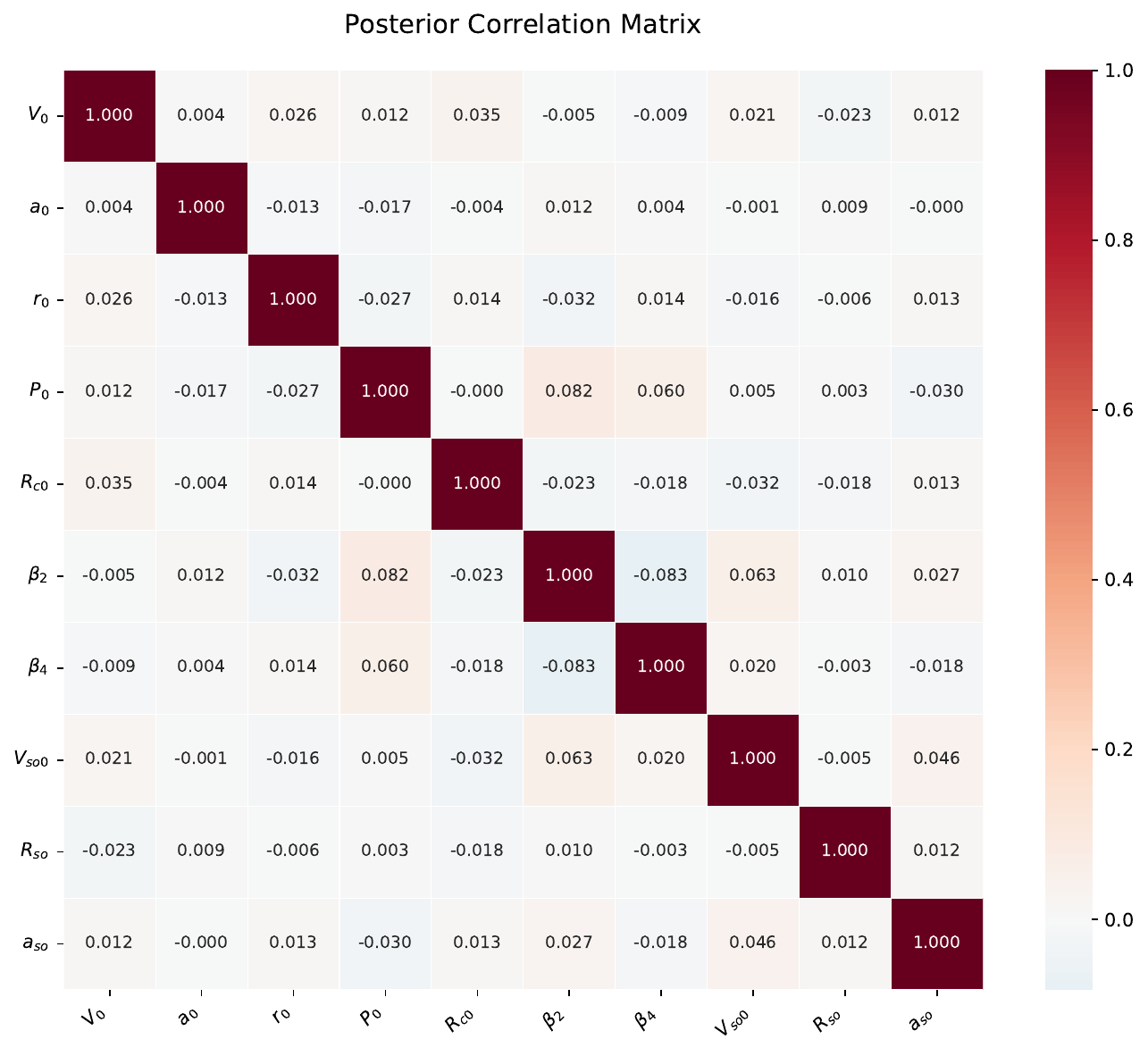}}
  \end{minipage}
  \caption{Supplementary posterior diagnostics for \nuc{121}{Pr}. Left: marginal and joint posterior distributions of the deformed-potential parameters obtained from MCMC sampling. Right: posterior Pearson correlation matrix of the sampled parameters, with the color scale indicating the linear correlation coefficient for each parameter pair.}
  \label{fig:app_diagnostics_121Pr}
  \label{fig:app_posterior_121Pr}
  \label{fig:app_correlation_121Pr}
\end{figure*}

\begin{figure*}[tbp]
  \centering
  \begin{minipage}[t]{0.49\textwidth}
    \centering
    \includegraphics[width=\linewidth]{{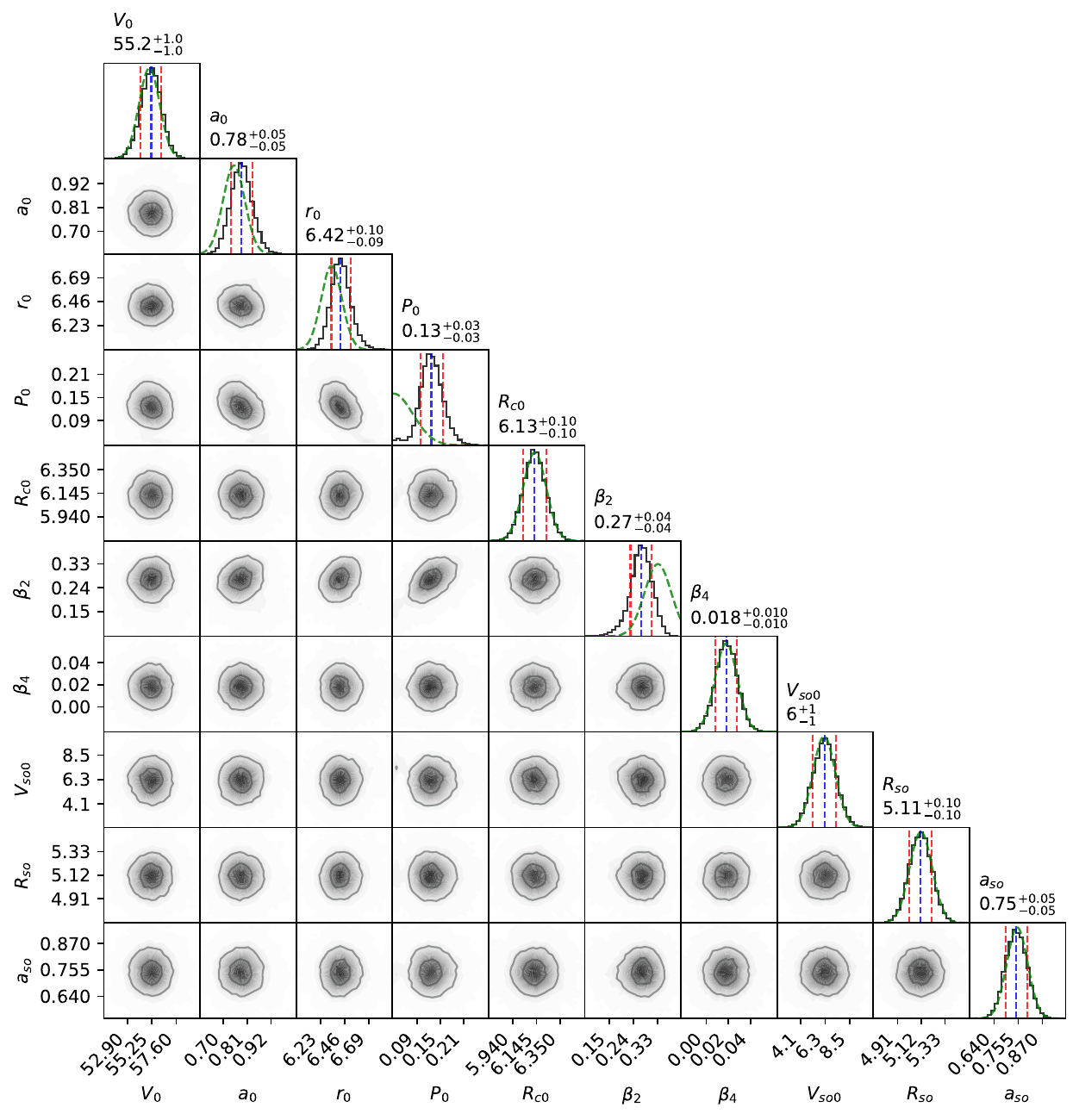}}
  \end{minipage}\hfill
  \begin{minipage}[t]{0.49\textwidth}
    \centering
    \includegraphics[width=\linewidth]{{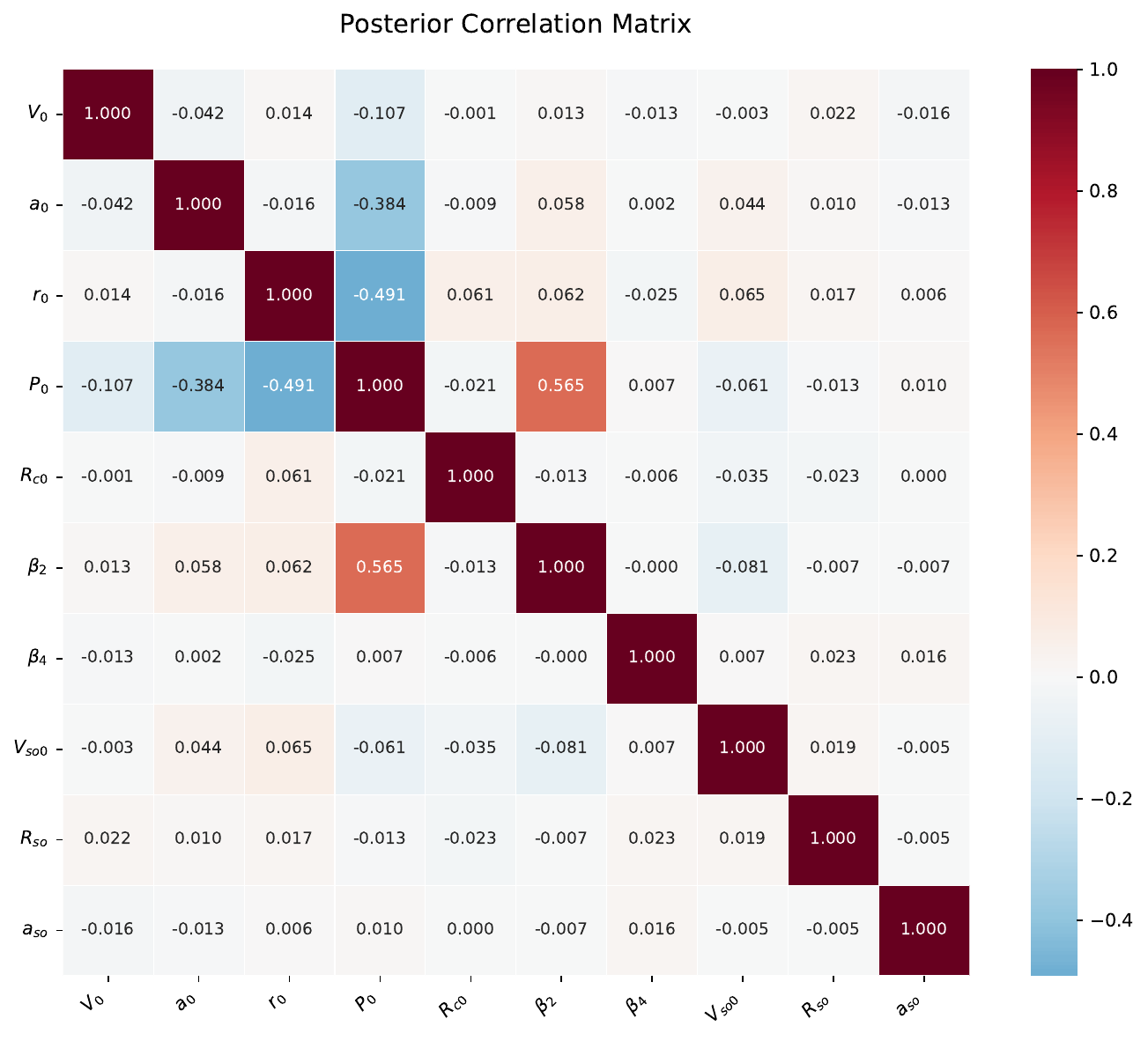}}
  \end{minipage}
  \caption{Supplementary posterior diagnostics for \nuc{131}{Eu}. Left: marginal and joint posterior distributions of the deformed-potential parameters obtained from MCMC sampling. Right: posterior Pearson correlation matrix of the sampled parameters, with the color scale indicating the linear correlation coefficient for each parameter pair.}
  \label{fig:app_diagnostics_131Eu}
  \label{fig:app_posterior_131Eu}
  \label{fig:app_correlation_131Eu}
\end{figure*}

\begin{figure*}[tbp]
  \centering
  \begin{minipage}[t]{0.49\textwidth}
    \centering
    \includegraphics[width=\linewidth]{{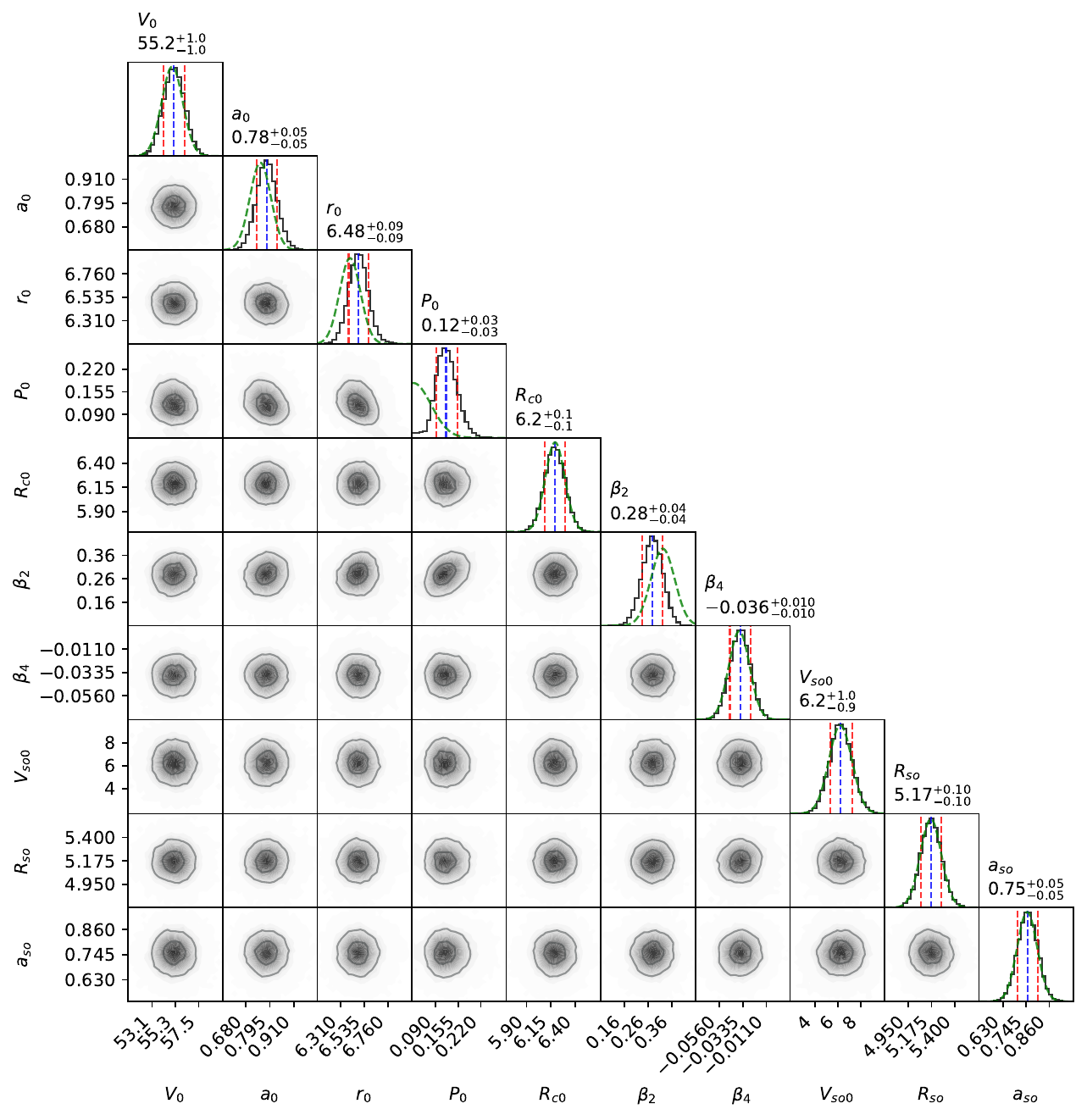}}
  \end{minipage}\hfill
  \begin{minipage}[t]{0.49\textwidth}
    \centering
    \includegraphics[width=\linewidth]{{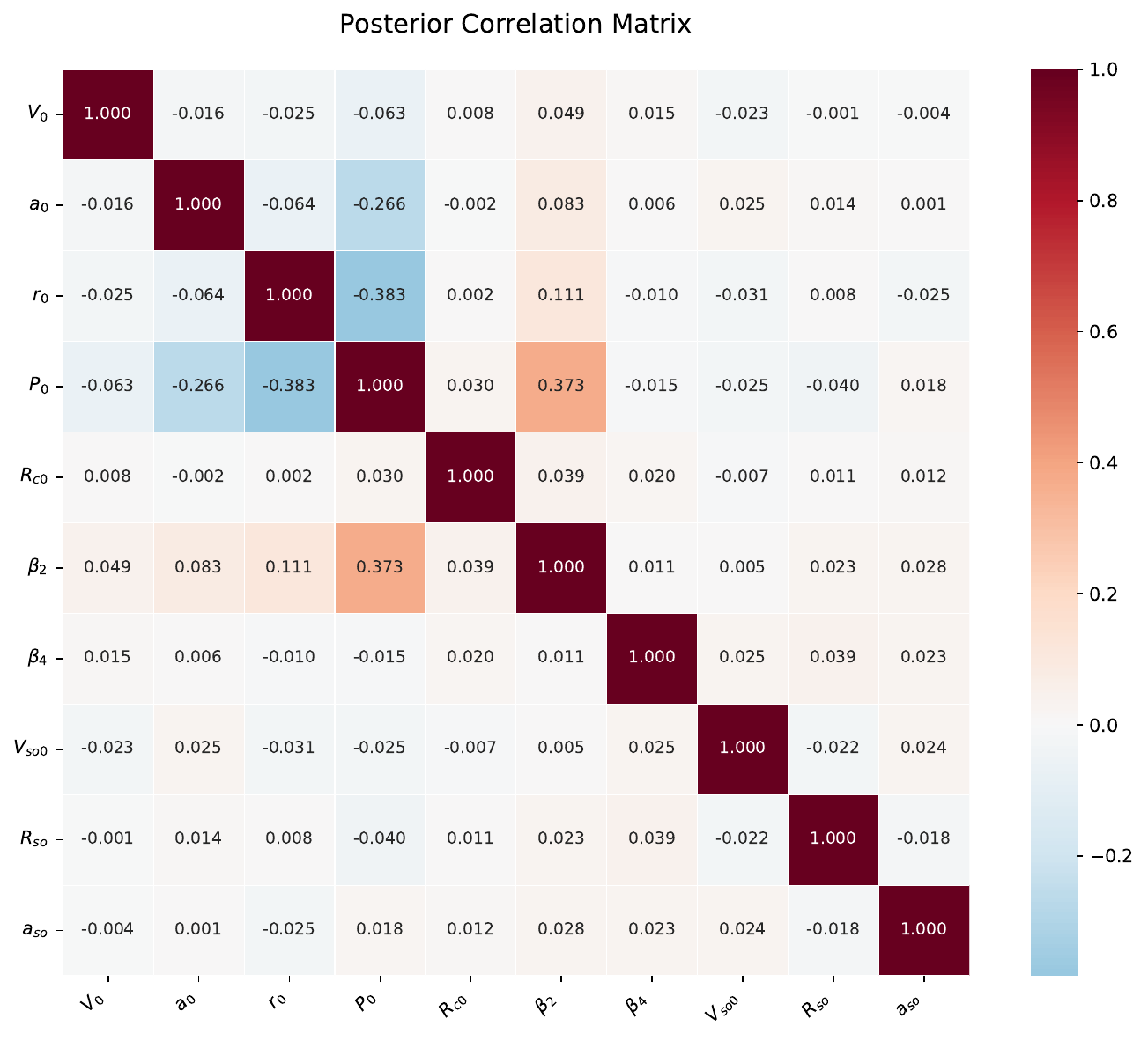}}
  \end{minipage}
  \caption{Supplementary posterior diagnostics for \nuc{135}{Tb}. Left: marginal and joint posterior distributions of the deformed-potential parameters obtained from MCMC sampling. Right: posterior Pearson correlation matrix of the sampled parameters, with the color scale indicating the linear correlation coefficient for each parameter pair.}
  \label{fig:app_diagnostics_135Tb}
  \label{fig:app_posterior_135Tb}
  \label{fig:app_correlation_135Tb}
\end{figure*}

\begin{figure*}[tbp]
  \centering
  \begin{minipage}[t]{0.49\textwidth}
    \centering
    \includegraphics[width=\linewidth]{{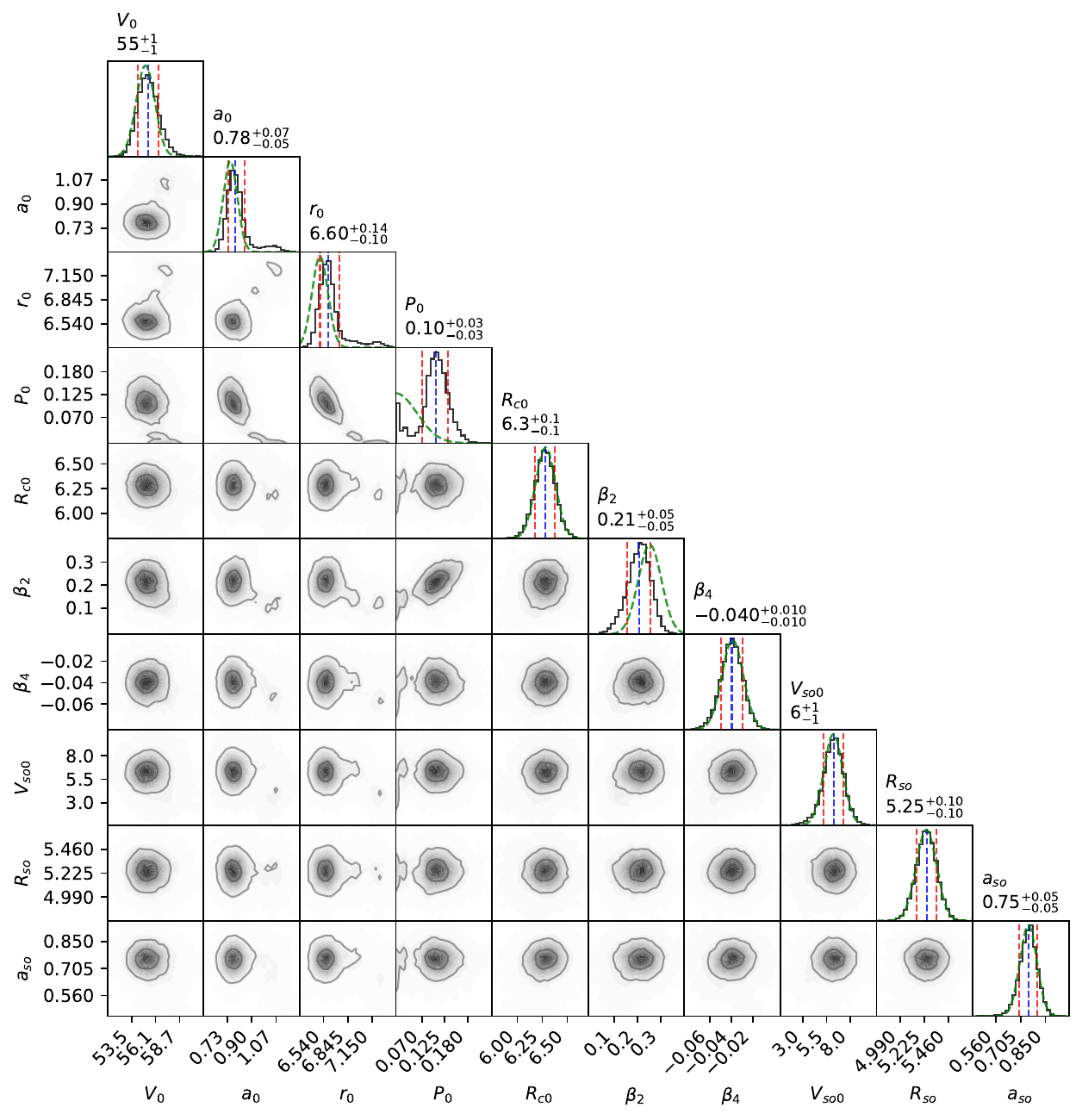}}
  \end{minipage}\hfill
  \begin{minipage}[t]{0.49\textwidth}
    \centering
    \includegraphics[width=\linewidth]{{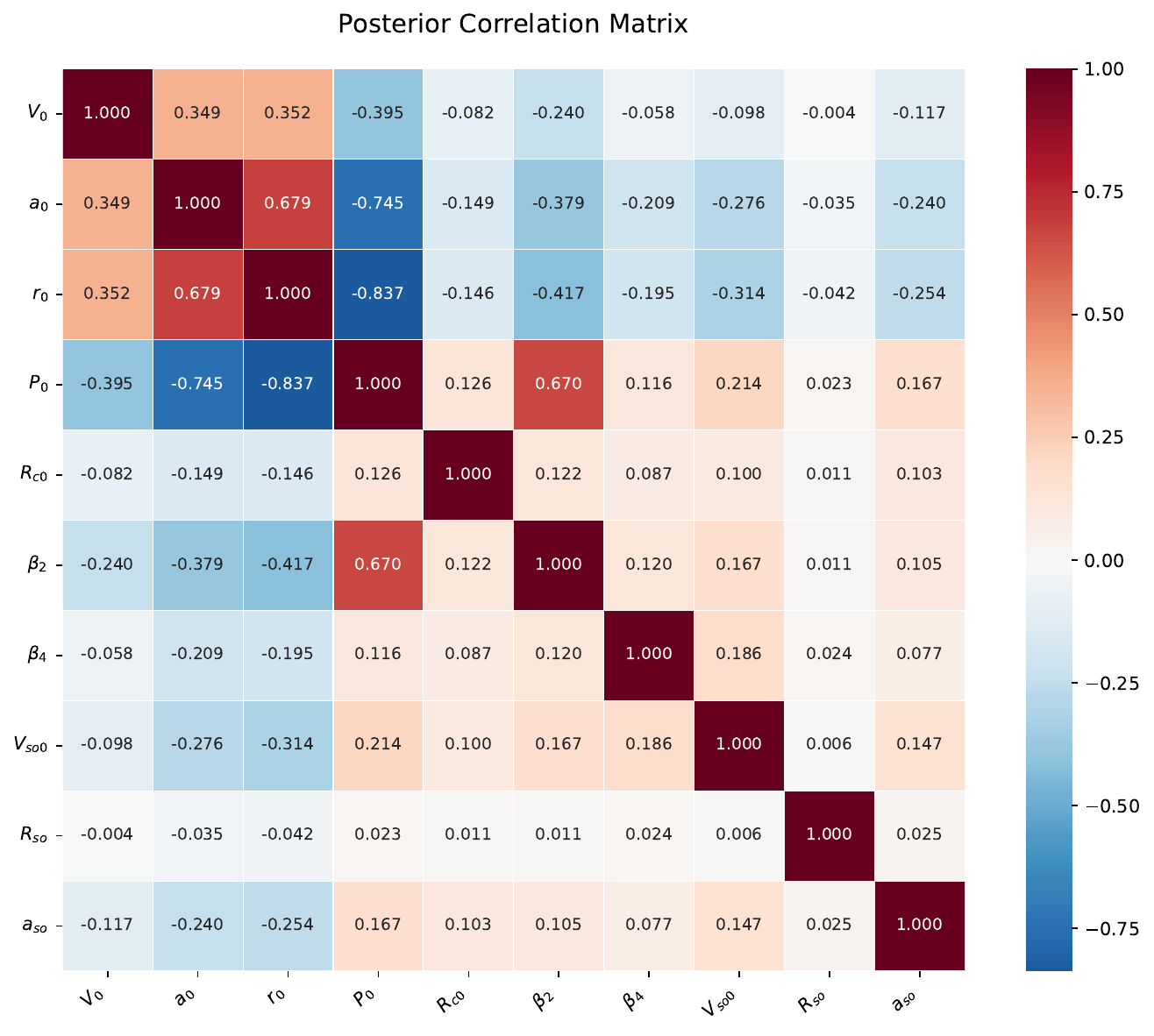}}
  \end{minipage}
  \caption{Supplementary posterior diagnostics for \nuc{141}{Ho}. Left: marginal and joint posterior distributions of the deformed-potential parameters obtained from MCMC sampling. Right: posterior Pearson correlation matrix of the sampled parameters, with the color scale indicating the linear correlation coefficient for each parameter pair.}
  \label{fig:app_diagnostics_141Ho}
  \label{fig:app_posterior_141Ho}
  \label{fig:app_correlation_141Ho}
\end{figure*}

\begin{figure*}[tbp]
  \centering
  \begin{minipage}[t]{0.49\textwidth}
    \centering
    \includegraphics[width=\linewidth]{{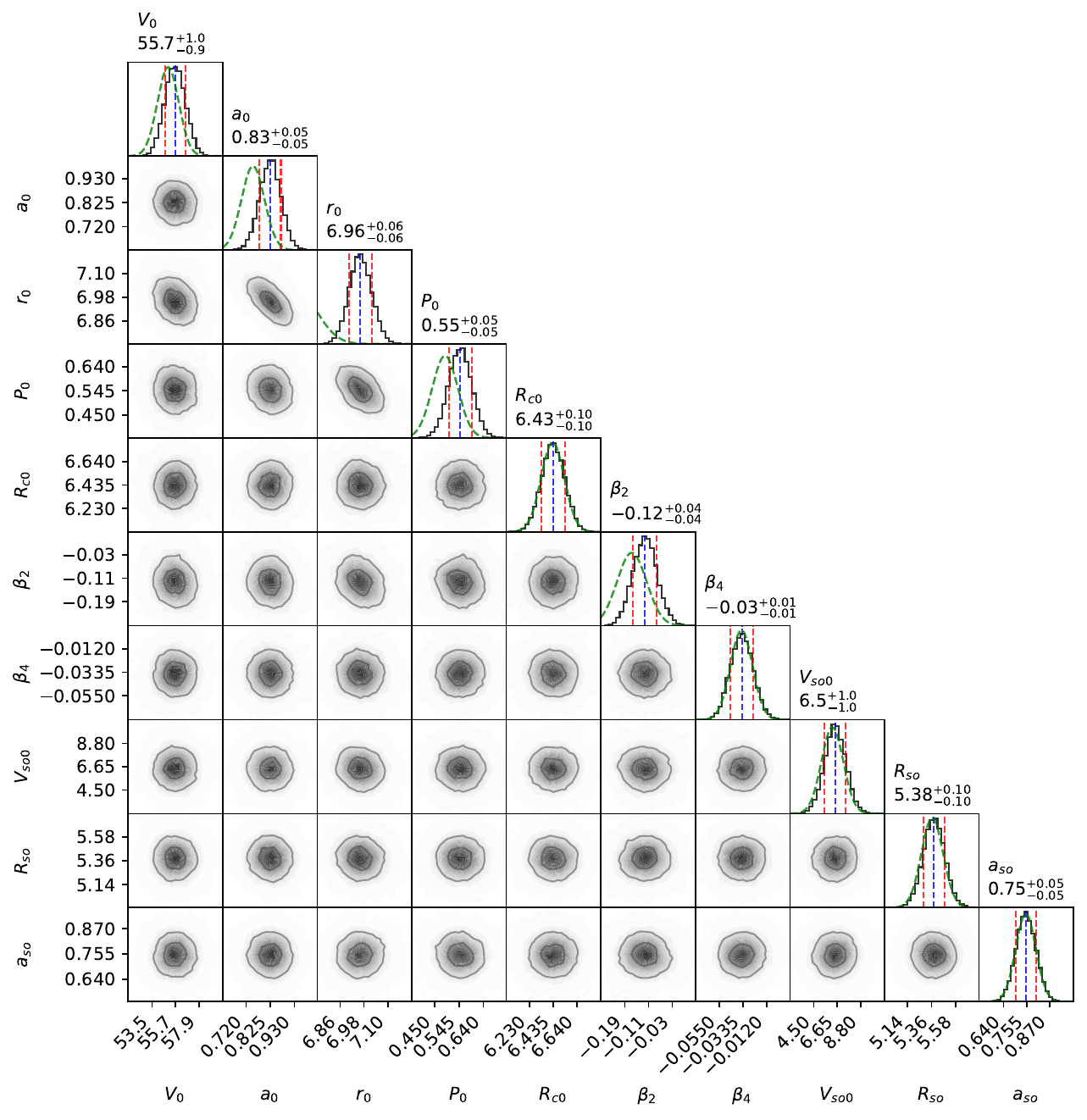}}
  \end{minipage}\hfill
  \begin{minipage}[t]{0.49\textwidth}
    \centering
    \includegraphics[width=\linewidth]{{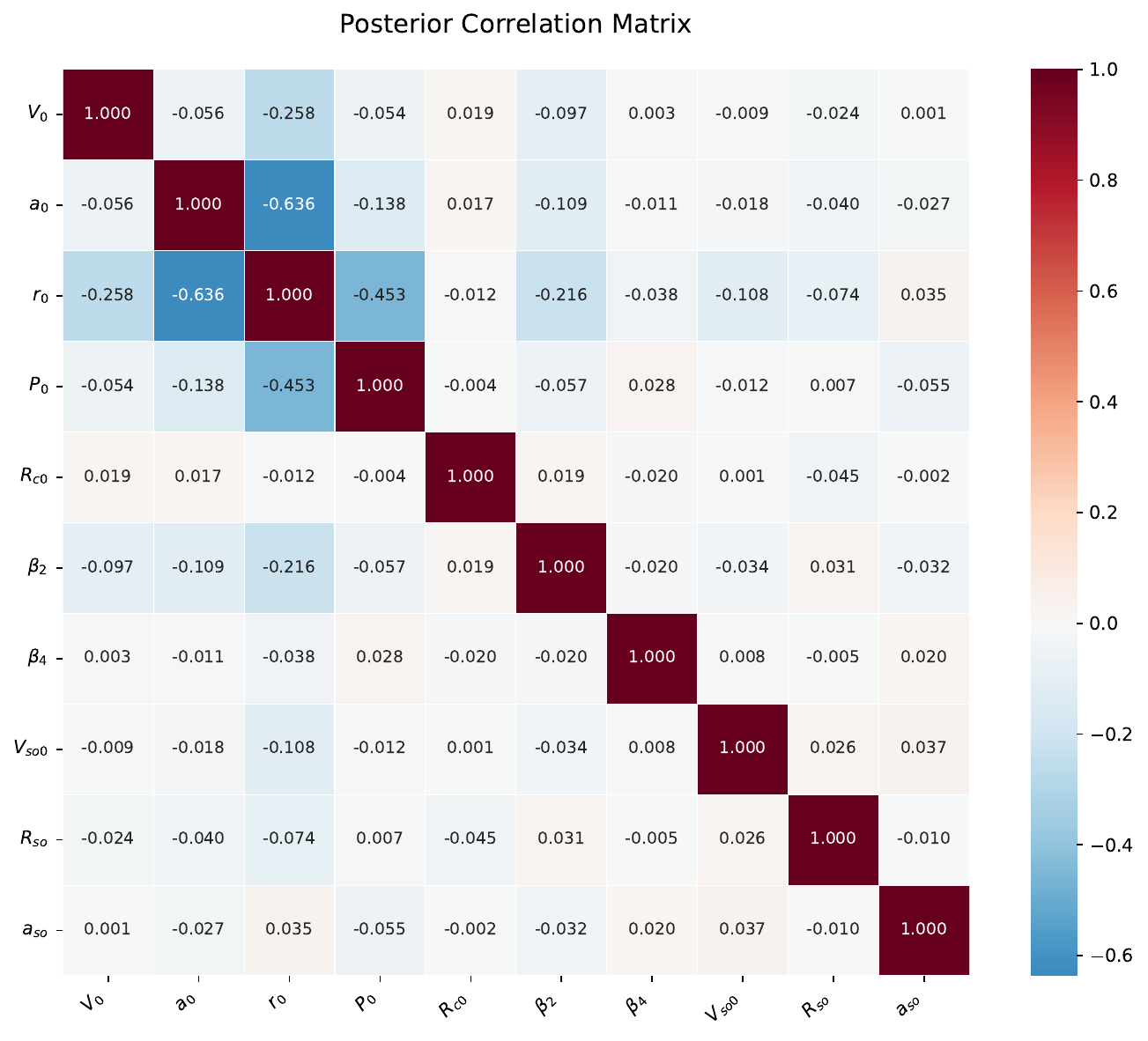}}
  \end{minipage}
  \caption{Supplementary posterior diagnostics for \nuc{151}{Lu}. Left: marginal and joint posterior distributions of the deformed-potential parameters obtained from MCMC sampling. Right: posterior Pearson correlation matrix of the sampled parameters, with the color scale indicating the linear correlation coefficient for each parameter pair.}
  \label{fig:app_diagnostics_151Lu}
  \label{fig:app_posterior_151Lu}
  \label{fig:app_correlation_151Lu}
\end{figure*}

\begin{figure*}[tbp]
  \centering
  \begin{minipage}[t]{0.49\textwidth}
    \centering
    \includegraphics[width=\linewidth]{{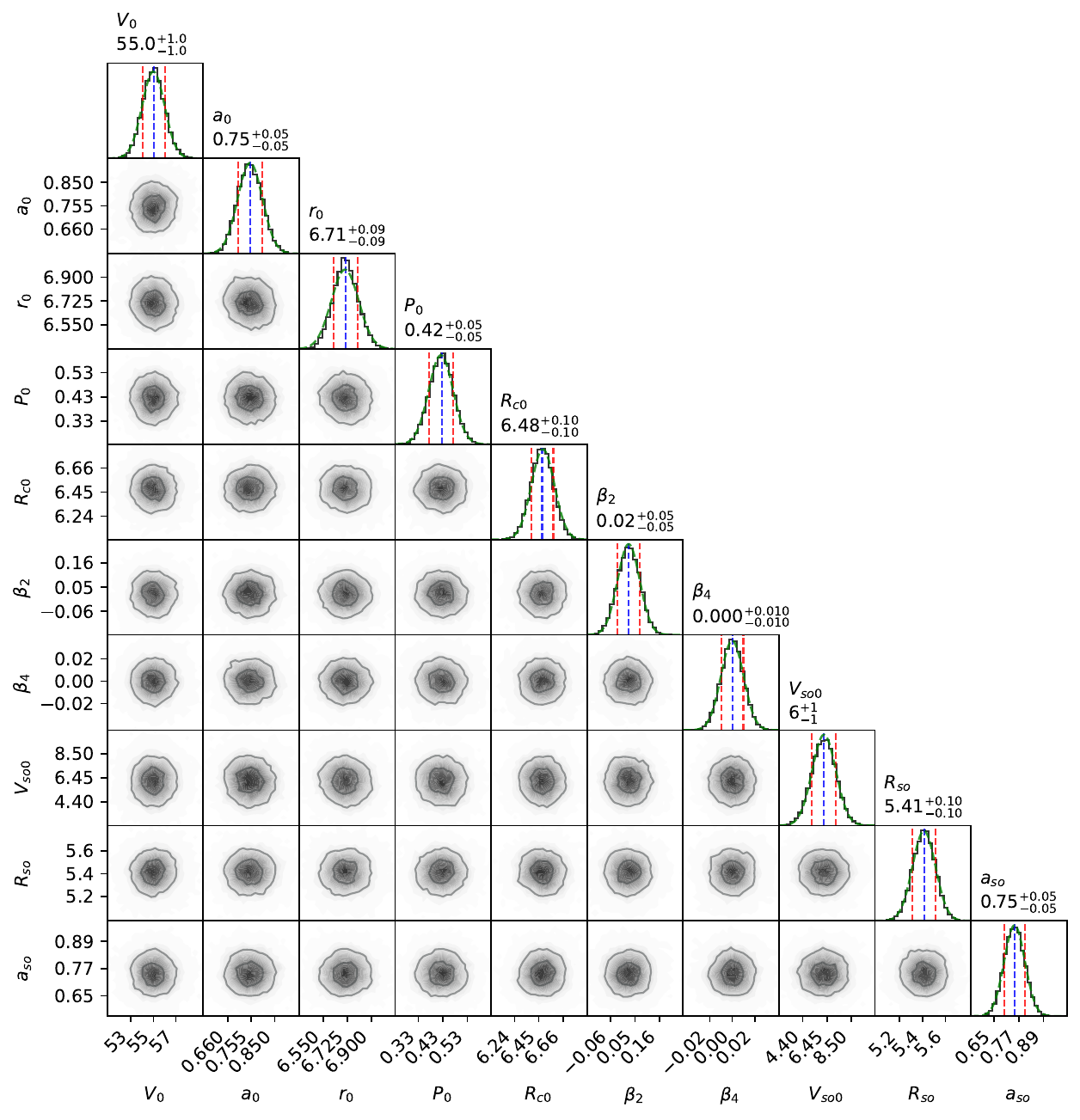}}
  \end{minipage}\hfill
  \begin{minipage}[t]{0.49\textwidth}
    \centering
    \includegraphics[width=\linewidth]{{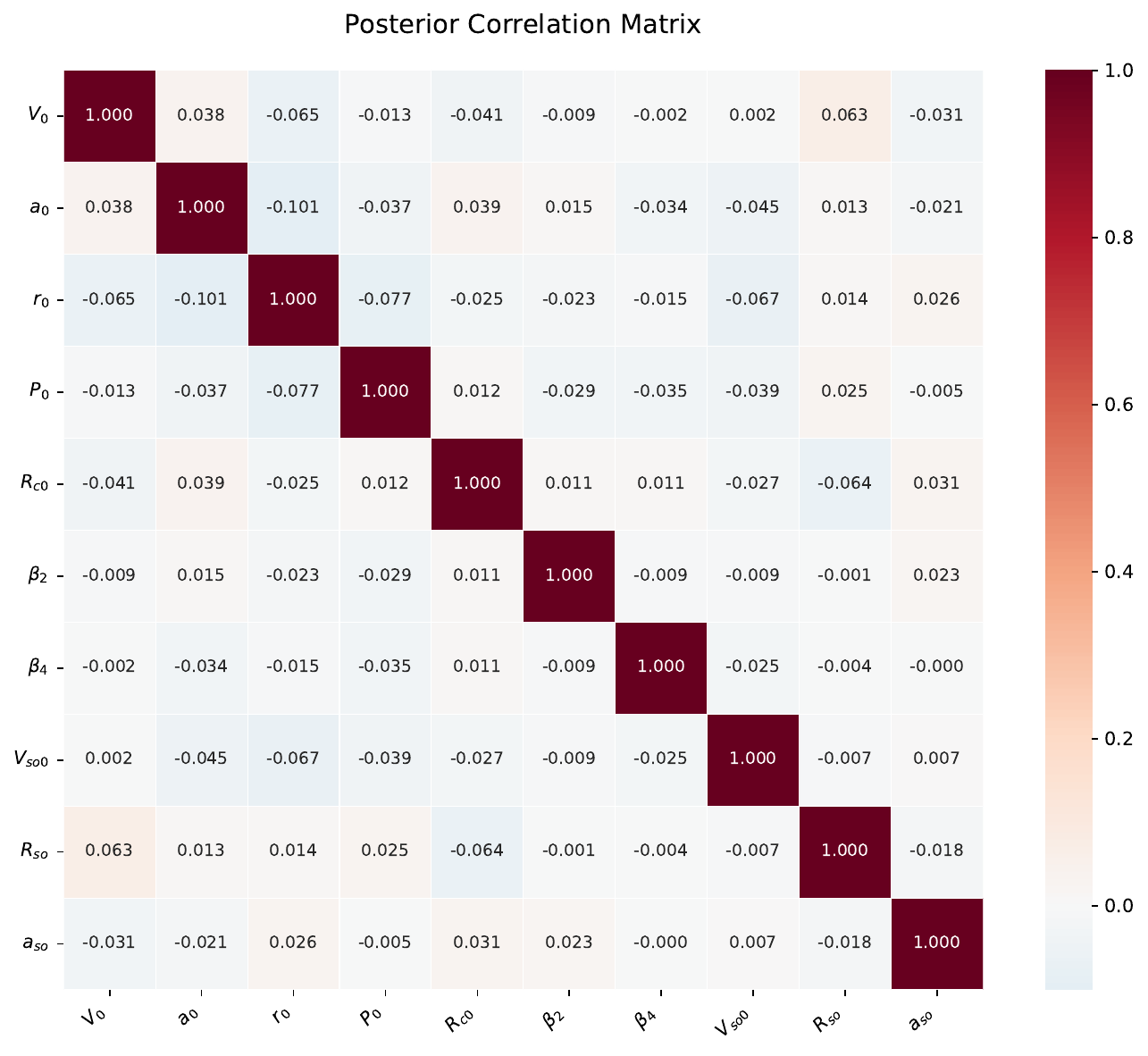}}
  \end{minipage}
  \caption{Supplementary posterior diagnostics for \nuc{155}{Ta}. Left: marginal and joint posterior distributions of the deformed-potential parameters obtained from MCMC sampling. Right: posterior Pearson correlation matrix of the sampled parameters, with the color scale indicating the linear correlation coefficient for each parameter pair.}
  \label{fig:app_diagnostics_155Ta}
  \label{fig:app_posterior_155Ta}
  \label{fig:app_correlation_155Ta}
\end{figure*}

\begin{figure*}[tbp]
  \centering
  \begin{minipage}[t]{0.49\textwidth}
    \centering
    \includegraphics[width=\linewidth]{{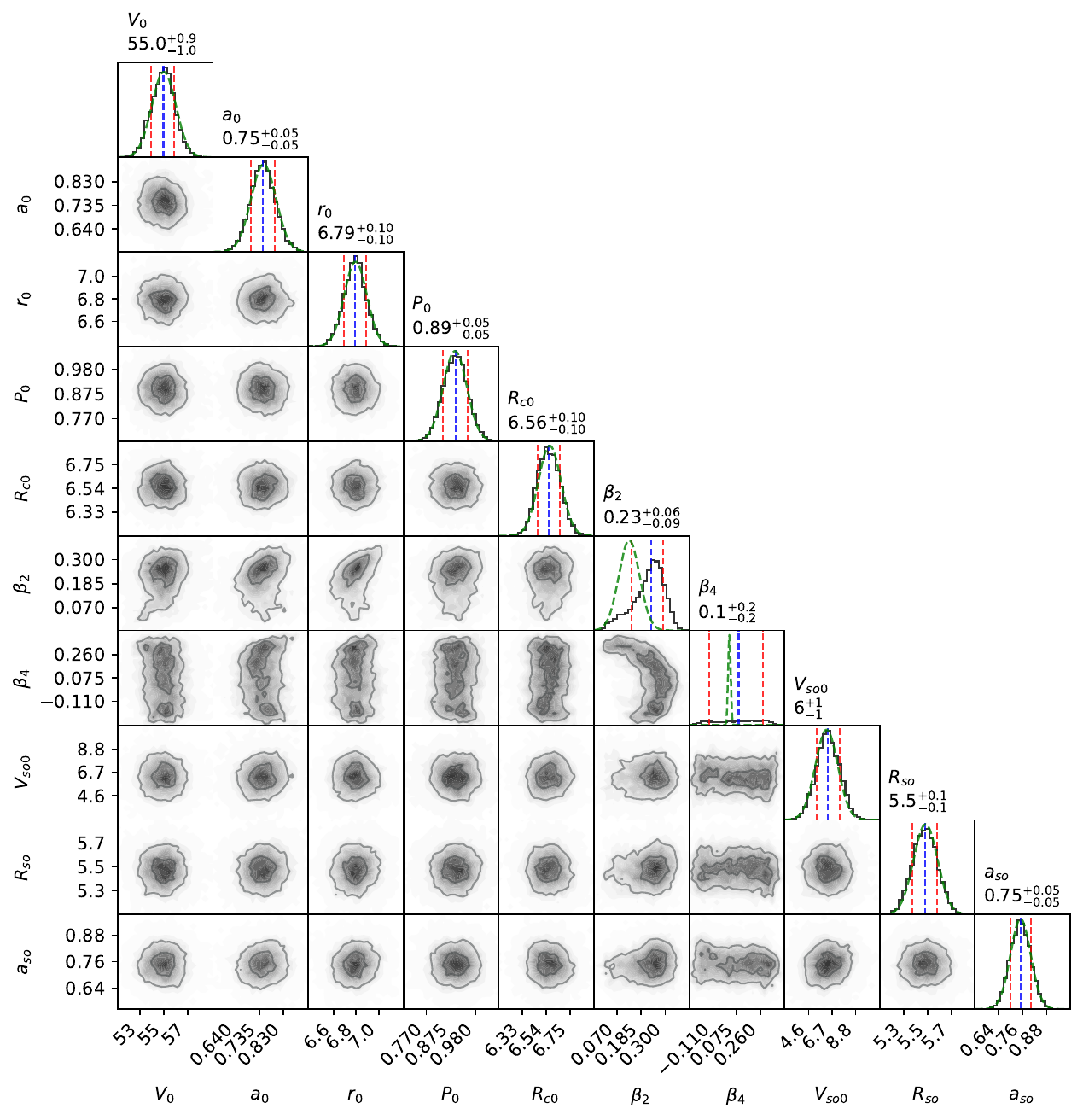}}
  \end{minipage}\hfill
  \begin{minipage}[t]{0.49\textwidth}
    \centering
    \includegraphics[width=\linewidth]{{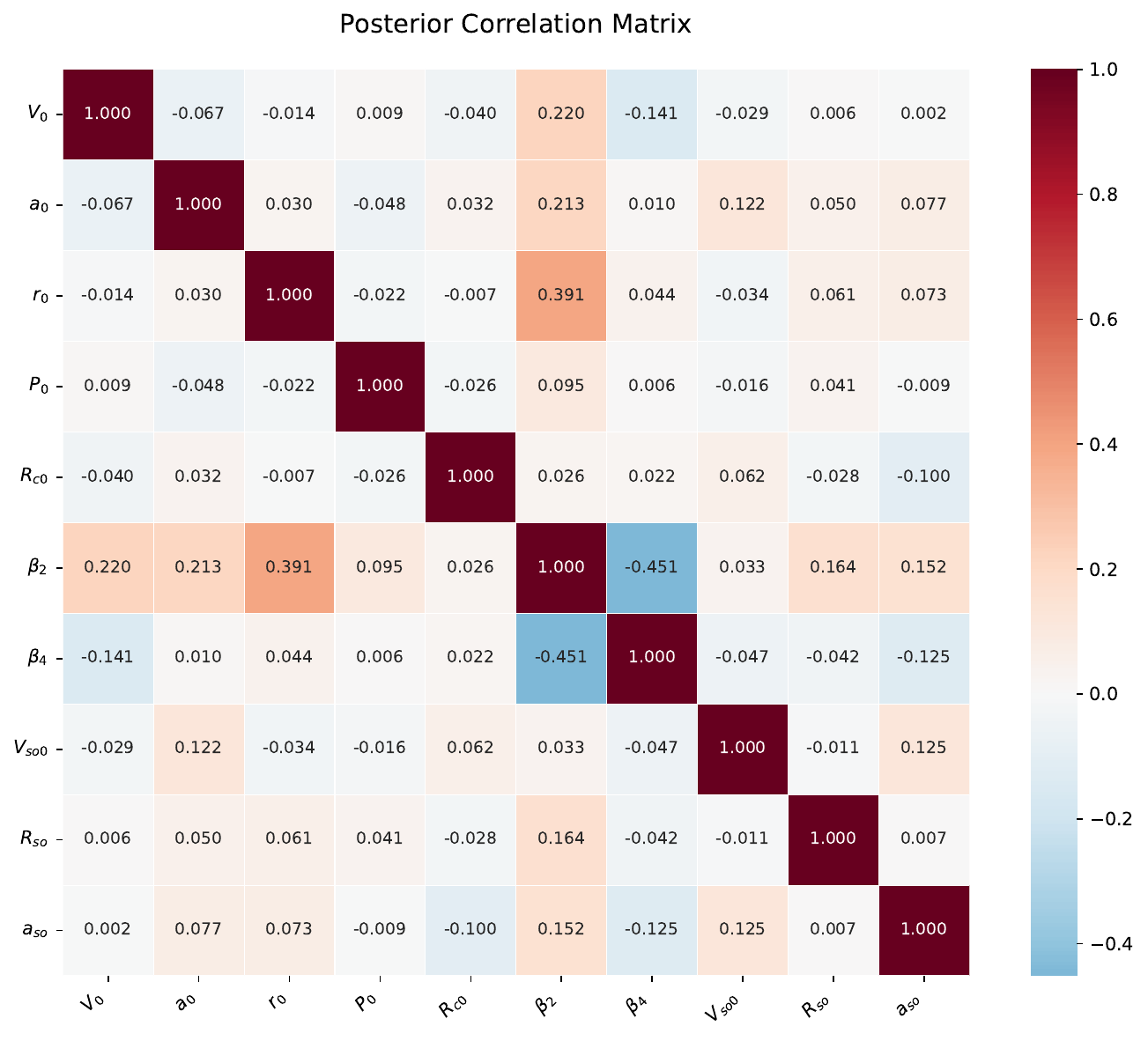}}
  \end{minipage}
  \caption{Supplementary posterior diagnostics for \nuc{161}{Re}. Left: marginal and joint posterior distributions of the deformed-potential parameters obtained from MCMC sampling. Right: posterior Pearson correlation matrix of the sampled parameters, with the color scale indicating the linear correlation coefficient for each parameter pair.}
  \label{fig:app_diagnostics_161Re}
  \label{fig:app_posterior_161Re}
  \label{fig:app_correlation_161Re}
\end{figure*}

\begin{figure*}[tbp]
  \centering
  \begin{minipage}[t]{0.49\textwidth}
    \centering
    \includegraphics[width=\linewidth]{{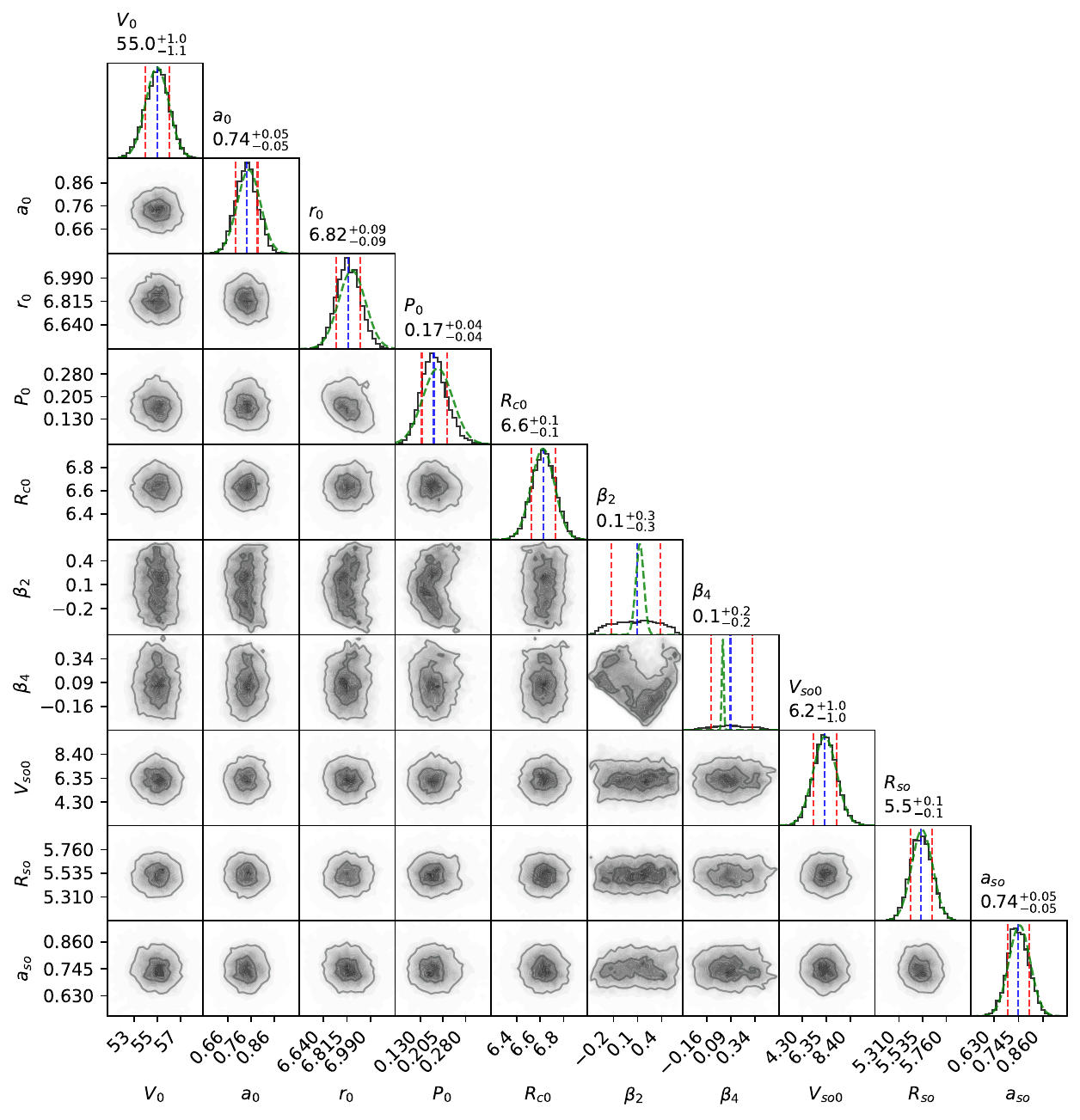}}
  \end{minipage}\hfill
  \begin{minipage}[t]{0.49\textwidth}
    \centering
    \includegraphics[width=\linewidth]{{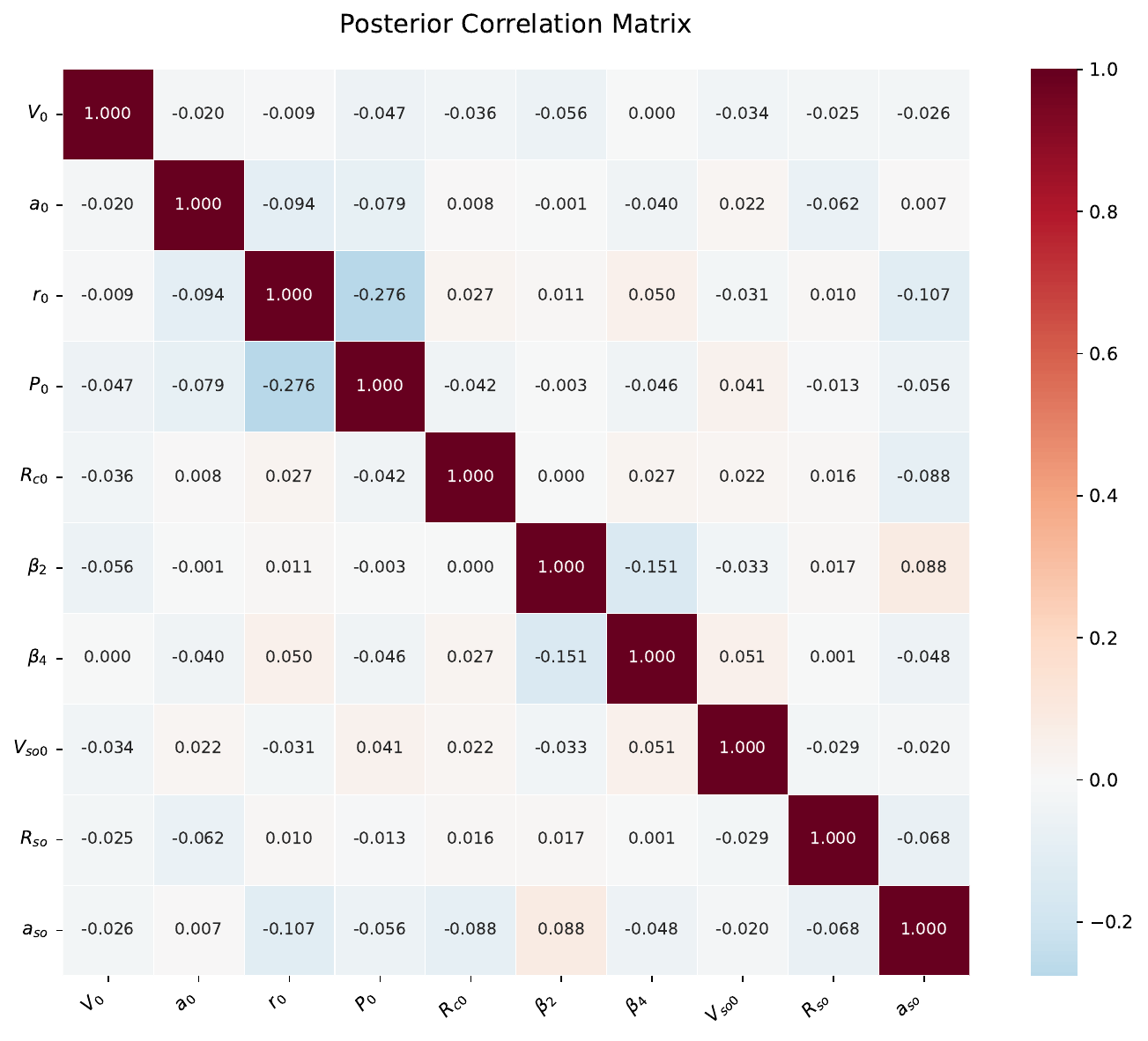}}
  \end{minipage}
  \caption{Supplementary posterior diagnostics for \ensuremath{{}^{165\mathrm{m}}\mathrm{Ir}}. Left: marginal and joint posterior distributions of the deformed-potential parameters obtained from MCMC sampling. Right: posterior Pearson correlation matrix of the sampled parameters, with the color scale indicating the linear correlation coefficient for each parameter pair.}
  \label{fig:app_diagnostics_165Ir_m}
  \label{fig:app_posterior_165Ir_m}
  \label{fig:app_correlation_165Ir_m}
\end{figure*}

\begin{figure*}[tbp]
  \centering
  \begin{minipage}[t]{0.49\textwidth}
    \centering
    \includegraphics[width=\linewidth]{{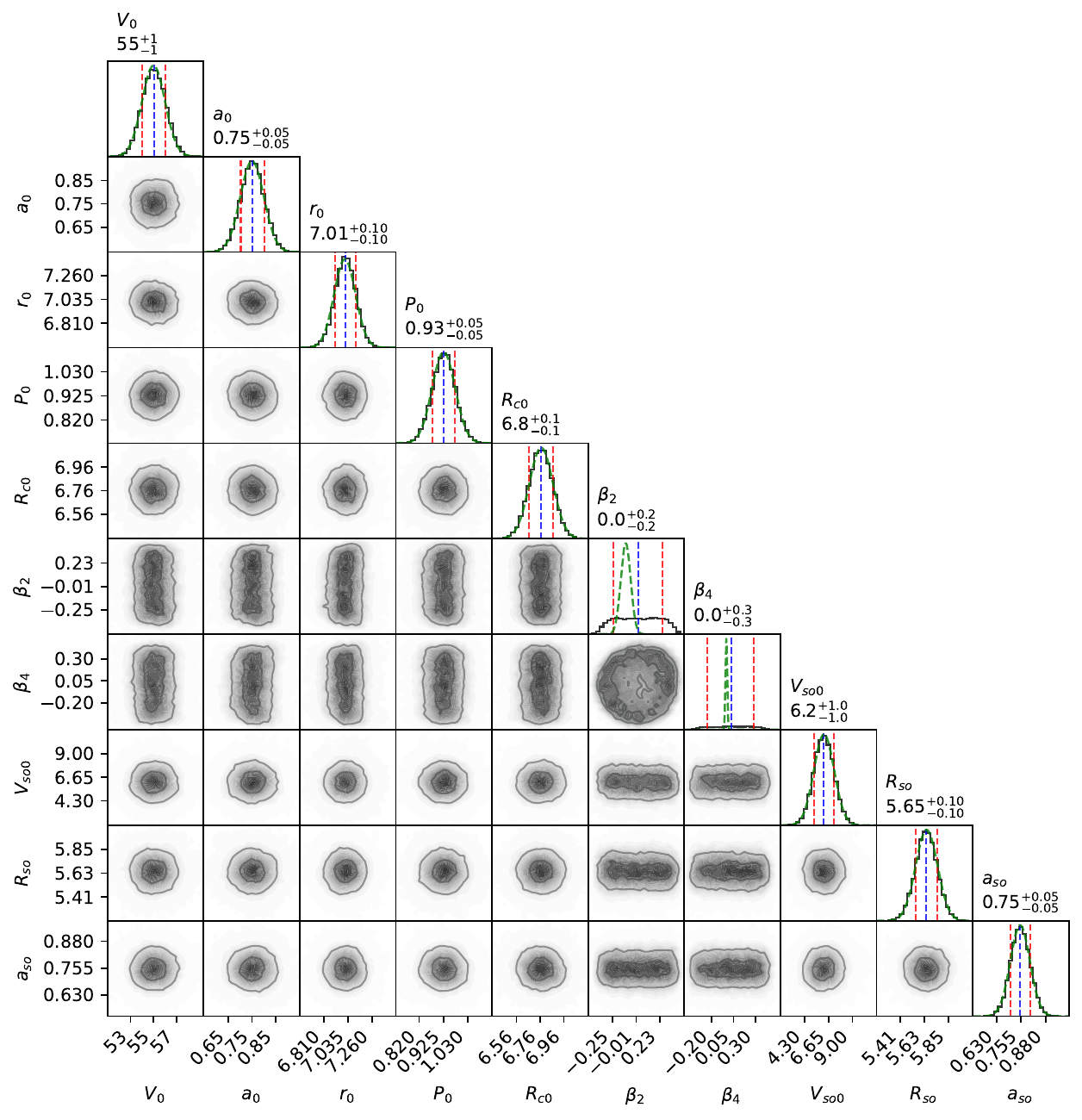}}
  \end{minipage}\hfill
  \begin{minipage}[t]{0.49\textwidth}
    \centering
    \includegraphics[width=\linewidth]{{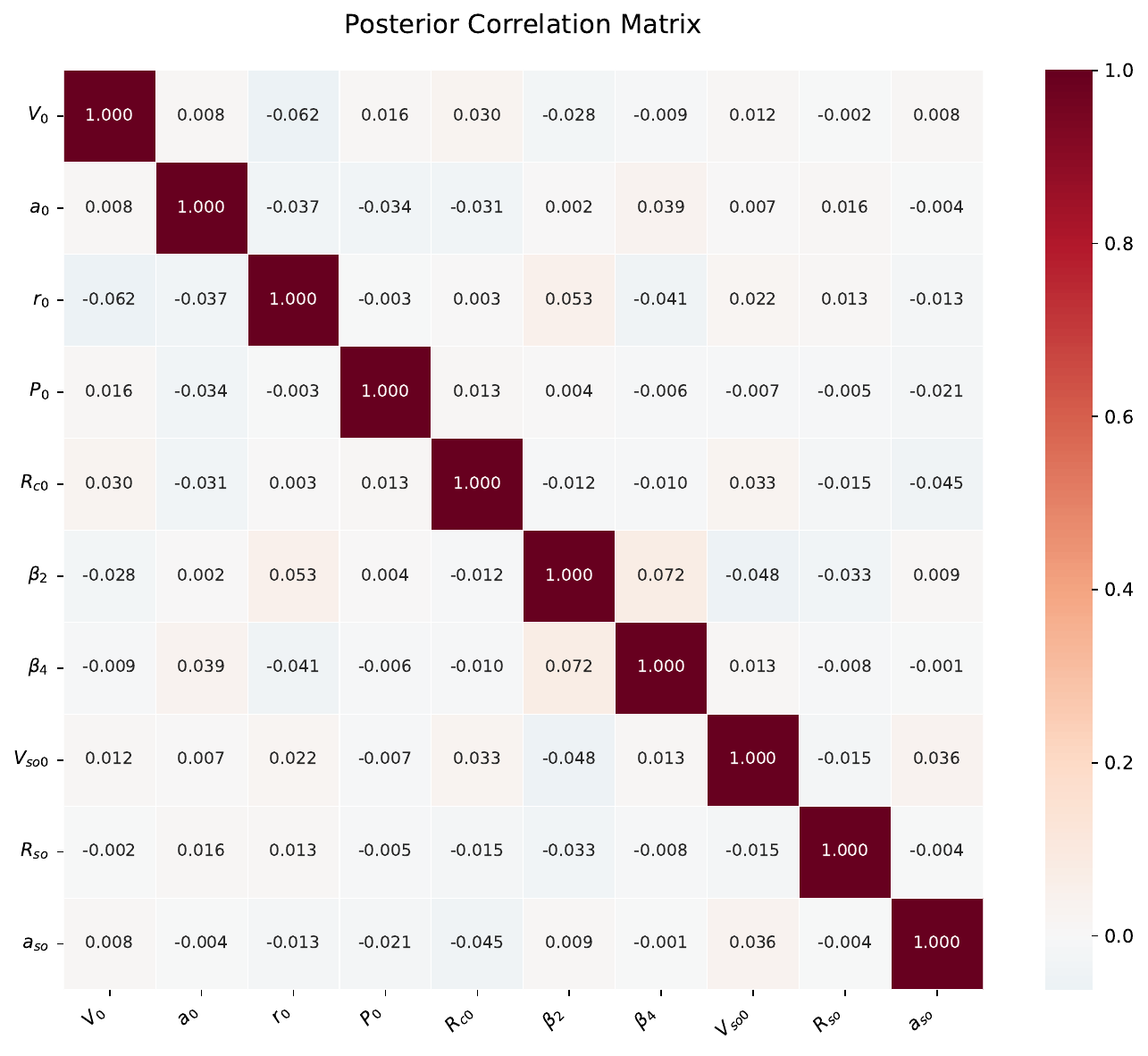}}
  \end{minipage}
  \caption{Supplementary posterior diagnostics for \nuc{176}{Tl}. Left: marginal and joint posterior distributions of the deformed-potential parameters obtained from MCMC sampling. Right: posterior Pearson correlation matrix of the sampled parameters, with the color scale indicating the linear correlation coefficient for each parameter pair.}
  \label{fig:app_diagnostics_176Tl}
  \label{fig:app_posterior_176Tl}
  \label{fig:app_correlation_176Tl}
\end{figure*}

\bibliography{apssamp}

\end{document}